\documentclass[onecolumn]{els-mrw_modified} 

\usepackage{amsmath,amssymb,amsfonts,amsthm,makeidx,graphicx}
\usepackage{txfonts}
\usepackage{helvet}

\newtheorem{theorem}{Theorem}
\newtheorem{conj}{Conjecture}
\newtheorem{definition}{Definition}

\newtheorem{rem}{Remark}

\newcommand{\cF}{\mathcal{F}}
\newcommand{\cO}{\mathcal{O}}
\newcommand{\cP}{\mathcal{P}}
\newcommand{\cR}{\mathcal{R}}
\newcommand{\cS}{\mathcal{S}}\newcommand{\cV}{\mathcal{V}}
\newcommand{\IN}{\mathbb{N}}
\newcommand{\IR}{\mathbb{R}}

\newcommand{\IE}{\mathbb{E}}
\newcommand{\IT}{\mathbb{T}}
\newcommand{\IZ}{\mathbb{Z}}
\newcommand{\balpha}{\boldsymbol{\alpha}}

\newcommand{\oOmega}{\mathring{\Omega}}
\newcommand{\tlambda}{\tilde\lambda}
\newcommand{\tk}{\tilde k}
\newcommand{\tchi}{\tilde \chi}
\newcommand{\tJ}{\tilde J}\newcommand{\vareps}{\varepsilon}
\newcommand{\eps}{\epsilon}

\def\bbbone{{\mathchoice {1\mskip-4mu {\rm{l}}} {1\mskip-4mu {\rm{l}}}
{ 1\mskip-4.5mu {\rm{l}}} { 1\mskip-5mu {\rm{l}}}}}

\newcommand{\defeq}{\stackrel{\rm{def}}{=}}
\newcommand{\la}{\langle}
\newcommand{\ra}{\rangle}
\newcommand{\supp}{{\rm supp}}
\newcommand{\Op}{\operatorname{Op}}
\newcommand{\Oph}{\operatorname{Op}_\hbar}
\newcommand{\vol}{\operatorname{vol}}
\newcommand{\spec}{\operatorname{spec}}
\newcommand{\tr}{\operatorname{tr}}
\newcommand{\Var}{\operatorname{Var}}
\newcommand{\tw}{\textwidth}

\begin{document}

\chapter{Quantum ergodicity and semiclassical measures: mathematical results}\label{chap1}

\author[1]{St\'ephane Nonnenmacher}%


\address[1]{\orgname{Laboratoire de Math\'ematiques d'Orsay}, \orgdiv{Universit\'e Paris-Saclay, CNRS}, \orgaddress{Rue Michel Magat, 91405 Orsay, France}}


\maketitle

\begin{glossary}[Keywords]
Quantum Chaos, Quantum Ergodicity, Delocalization, Semiclassical Analysis, Semiclassical Measures

\end{glossary}

\begin{abstract}[Abstract]
In this chapter we review some results describing the high-frequency eigenmodes of the Laplacian on compact manifolds, or Euclidean domains, for which the geodesic flow is chaotic. We focus on the macroscopic distribution of these eigenmodes, which is described by the concept of semiclassical measure. The main result on the question is the Quantum Ergodicity theorem, originally due to Schnirelman. We provide the detailed proof of this theorem, including the adjustments necessary to treat the case of manifolds with boundary. We also discuss the Quantum Unique Ergodicity conjecture, and some progress towards this conjecture for strongly chaotic (Anosov) systems. In particular, we describe the constraints on admissible semiclassical measures, in terms of their Kolmogorov-Sinai entropy, as well as more recent delocalization results. 
\end{abstract}

\section{Introduction}\label{s:intro}

Acoustic or electromagnetic waves propagating inside a ``cavity''
naturally lead to the concept of {\it stationary vibration modes}, which
decouple the time and space degrees of freedom in the wave
equation. Mathematically, those modes are the eigenmodes of the
Laplace operator in the cavity, subject to appropriate boundary
conditions, depending on the physics of the problem (nature of the
wave, and of the cavity boundaries). A bounded cavity $\Omega \subset
\IR^d$ (in practice the dimension will be $d=1,2,3$) with Dirichlet
boundary conditions leads to a discrete set of
stationary modes $(u_n)_{n\geq 1}$ in correspondence with
eigenfrequencies $(\lambda_n)_{n\geq 1}$, which satisfy Helmholtz's
equation:
\begin{equation}\label{e:Helm}
(\Delta + \lambda_n^2)u_n(x) = 0,\qquad u_{n|\partial\Omega}=0.
\end{equation}
The eigenfrequencies are taken in increasing order,
$\lambda_n\to\infty$ as $n\to\infty$ and there can be
multiplicities $\lambda_n=\lambda_{n+1}$. We normalize those eigenfunctions by
$\|u_n\|_{L^2(\Omega)}=1$.

This discrete spectrum allows one to solve the wave equation in the
cavity (with sound speed $c=1$),
$(\partial_t^2+ \Delta)u(t,x)=0$ through the eigenmode decomposition,
$$
u(t,x) = \sum_{n\geq 1}\big(\alpha_n \cos(t\lambda_n)+\beta_n\sin(t\lambda_n)\big)\,u_n(x)\,,
$$
the coefficients $\alpha_n,\beta_n$ being fixed through the information on the initial data $u(0,x)$, $\partial_t
u(0,x)$. 

Formally, this decomposition fully solves the wave equation. But in
practice, what do we know about the stationary modes $(u_n)_{n\geq
  1}$? 

The answer depends a lot on the shape of the cavity $\Omega$,
and its symmetries.

\subsection{Integrable shapes}
In standard textbooks in wave physics, one often
presents the situation for specific shapes, which enjoy particular
symmetries, for instance a rectangular cavity in 2 dimensions, $\cR=[0,L_1]\times[0,L_2]$, for
which one can separate the horizontal and vertical variables, and
solve the Helmholtz equation through the factorized Ansatz
$u_n(x_1,x_2)=v_n(x_1)w_n(x_2)$. Each factor is now solution of an
ordinary differential equation involving one term of the Laplacian:
\begin{align*}
  v_n''(x_1)+\lambda_{n,1}^2v_n(x_1)=0,&\quad v_n(0)=v_n(L_1)=0,\\
  w_n''(x_2)+\lambda_{n,2}^2w_n(x_2)=0,&\quad w_n(0)=w_n(L_2)=0,
\end{align*}
which are solved by sine functions $v_n(x_1)=\sin(\lambda_{n,1}x_1)$,
with the boundary constraints $\lambda_{n,i}\in \frac{\pi}{L_i}\IN$
for $i=1,2$.
Putting together these
1D solutions we recover a 2D eigenmode $u_n(x_1,x_2)=v_n(x_1)w_n(x_2)$
with eigenfrequency
$\lambda_n=\sqrt{\lambda_{n,1}^2+\lambda_{n,1}^2}$. Due to this
separation of variables, it rather makes sense to label the
eigenfunctions by the pair of integers
$m_i=\frac{\lambda_{n,i}L_i}{\pi}$, rather than a single integer $n$.

\begin{figure}[ht]
  \begin{center}
  \includegraphics[width=0.65\tw]{
  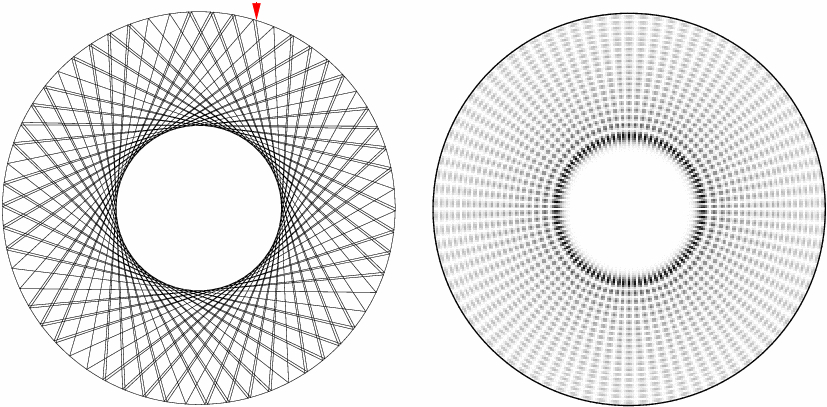}$\ $
  \includegraphics[width=0.33\tw]{
  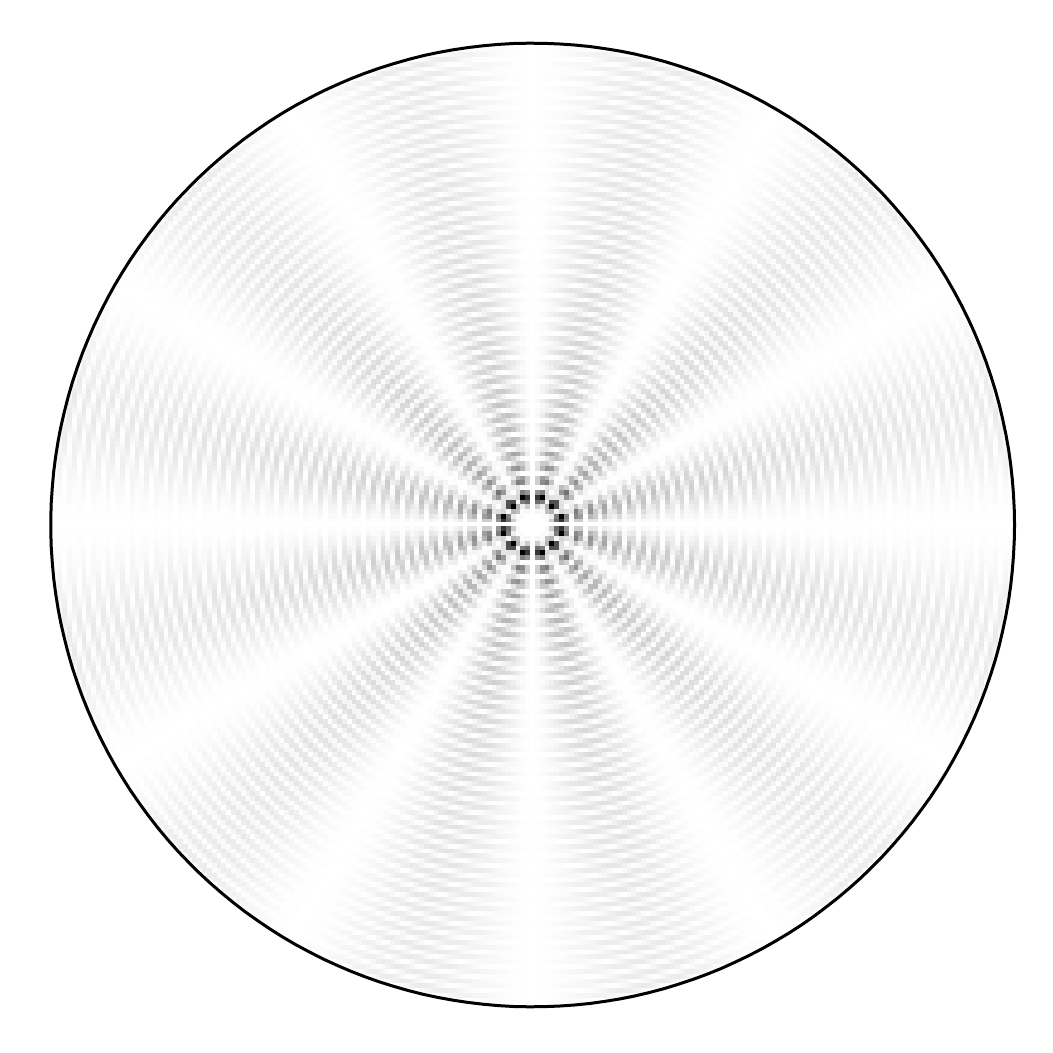}
  \caption{\label{f:circle}Left: one long ray inside the circle
    billiard (the red arrow is the starting point). Center: one
    eigenmode, with angular momentum $k$ equal to that of the ray (the
    grey hue corresponds to the intensity of $|u_n(x)|^2$). Right: eigenmode of the cirle with
    smaller angular momentum. Courtesy of Arnd B\"acker.}
  \end{center}
\end{figure}

A similar factorization can be obtained for a 2D cavity in the shape
of a disk centered at the origin of radius $R$. It then makes sense to use polar coordinates, in which we
can factorize the eigenmode through
\begin{equation}\label{e:disk-mode}
  u(r,\theta) = e^{ik\theta} J_{|k|}(\lambda r),
\end{equation}
where $k\in \IZ$ and $J_{|k|}$ is the 
Bessel function of order $|k|$. The eigenfrequency $\lambda$ is fixed
by the boundary condition $J_{|k|}(\lambda R)=0$. In
Fig.~\ref{f:circle} we plot two eigenstates of the circle, given by
the real parts states of the form
\eqref{e:disk-mode}, for different values of $k$ and $\lambda$.

In these two examples, the separation of variables is associated with
a continuous symmetry of the shape $\Omega$. This symmetry holds
at the ``quantum'' (or operator) level: for the rectangle, the Laplacian $\Delta_\Omega$ commutes with
the 1D Laplacians $\frac{\partial^2}{\partial x_1^2}$,
$\frac{\partial^2}{\partial x_1^2}$ equipped with separate boundary
conditions. In the case of the disk, the Laplacian commutes with the
angular momentum operator $-i\frac{\partial}{\partial\theta}$, which
is diagonalized by the modes \eqref{e:disk-mode}.

But these symmetries are already present and relevant when describing
the {\it ray dynamics} in $\Omega$, equivalent with the billiard
dynamics in $\Omega$. This dynamics describes the trajectory of points
in the phase space $\Omega\times\IR^2=\{(x,v),\ x\in\Omega,\
\xi\in \IR^2\}$ given by straight segments inside $\Omega$ followed by
specular reflections at the boundary.

In the case of the rectangle, if a particle
starts from $x\in\cR$ with velocity $v=(v_1,v_2)$, then not only is
the total velocity $|v|$ preserved by the flow (invariance of the
kinetic energy), but also the projections $|v_1|$ and $|v_2|$ are
preserved along the flow. In the case of the disk, the angular
momentum (w.r.t. the origin) $L=x\wedge v$ is also preserved along the
flow (when $|v|=1$, $|L|$ is equal to the distance of the ray from the
center of the disk).
These symmetries provide a very particular structure of the
billiard flow: the phase space $\Omega\times\IR^2$ is foliated by 2-dimensional
invariant tori determined by the conserved quantities: in the
rectangle, by the two projections of the velocity $(|v_1|,|v_2|)$, in the
circle by $(|v|, L)$.
Such billiard flows are said to be {\it integrable}
in the sense of Liouville, since they have as many conserved
quantities as degrees of freedom.

\subsection{Nonintegrable shapes: numerical observations}

The shapes $\Omega$ enjoying these symmetries are ``accidental'' among all
possible shapes. As a result, for a generic shape $\Omega$, there is
no way to separate variables and transform the Helmholtz equation into
a set of ODEs. In this situation, the Helmholtz equation is an
``irreducible'' PDE, and its solutions do not admit any simple
expression in terms of special functions. What can be done to gather
information on those eigenmodes?

One can solve try to solve (approximately) the Helmholtz equation
numerically, using for instance finite element methods inside
$\Omega$; instead one can decompose
the solution $u_n(x)$ into a sum of elementary functions (Fourier modes,
or Bessel functions), and impose the Dirichlet boundary condition to
fix the eigenvalues and eigenfunctions. Those methods works well at
low frequency, but
become computationally demanding when the frequency increases.
Indeed, at the frequency $\lambda_n$ the eigenmodes typically oscillate on the
scale $\sim\lambda_n^{-1}$, so when $\lambda_n\gg 1$ one needs fine
grids to keep a good precision on these oscillatory functions.
From the beginning of the 1980s, numerical computations of Laplacian
eigenmodes have been instrumental to ask new questions and discover
new phenomena, like the scarring of classical periodic orbits. The
fast increase of computer performances has allowed to reach relatively
high values of the frequency, at least in 2-dimensional problems.

\begin{figure}[ht]
  \begin{center}
\includegraphics[width=0.3\tw]{
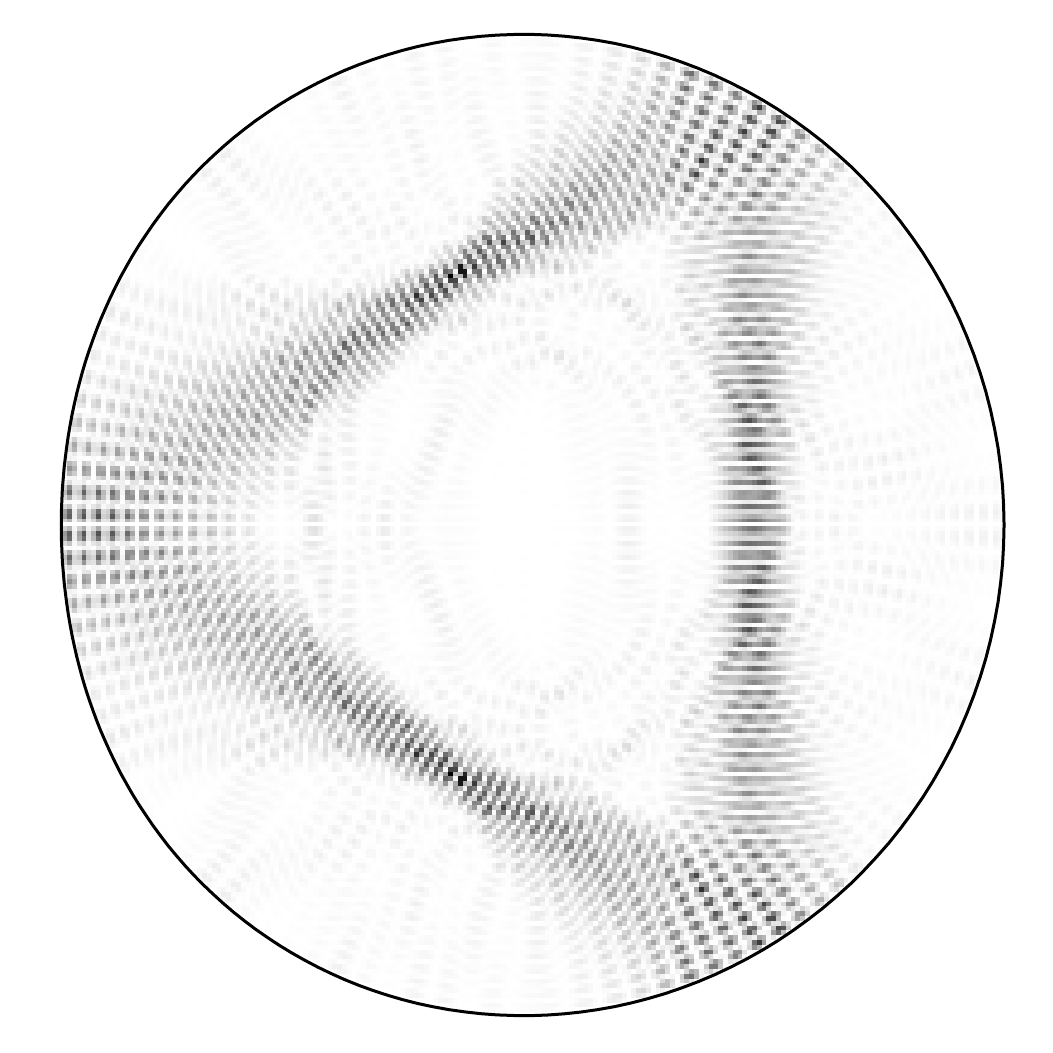} \hspace{.5cm}
\includegraphics[width=0.3\tw]{
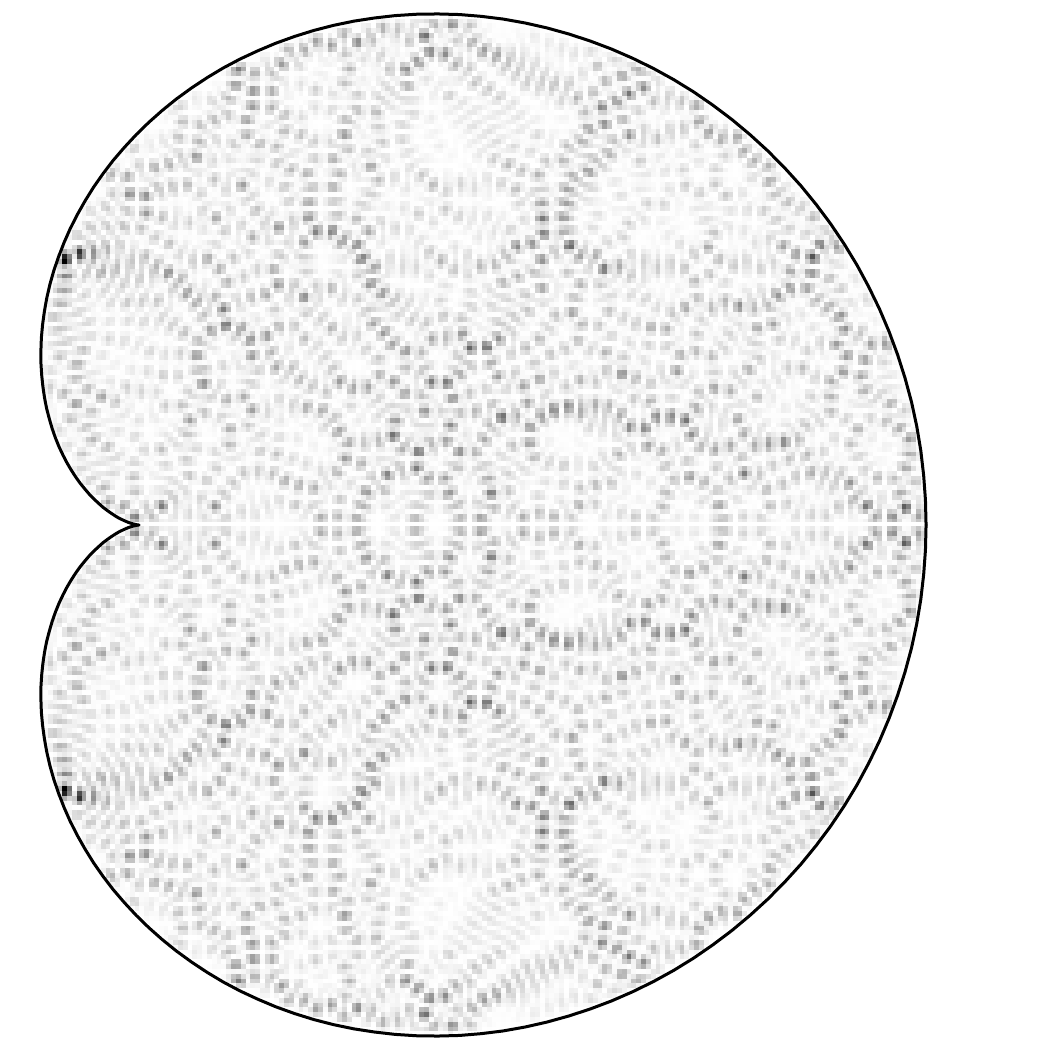} \hspace{.5cm}
\includegraphics[width=0.3\tw]{
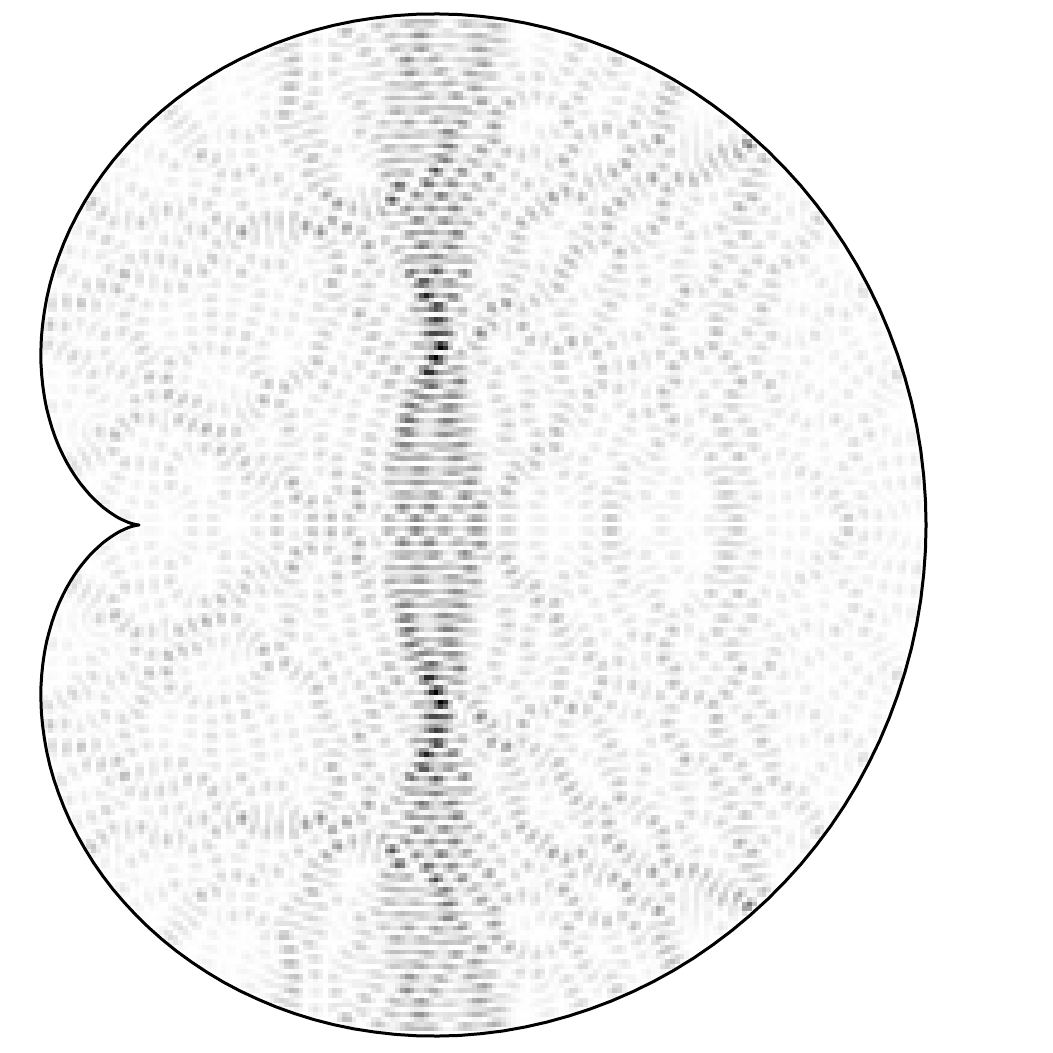}
\caption{\label{f:4modes}High frequency eigenmodes for a small deformation
  of the circle (left), and two eigenmodes of the cardioid. Courtesy of Arnd B\"acker.}
\end{center}
\end{figure}

In Figure~\ref{f:4modes} we represent several Dirichlet eigenmodes for
several shapes $\Omega$, at relatively high frequencies. On the left:
the lima\c con is a small deformation of the circle. The geodesic flow in
this lima\c con billiard encompasses some {\it stable} periodic
orbits, due to the KAM phenomenon. The shown eigenmode is
concentrated in the neighbourhood of one such short stable periodic
orbit. We also notice that the fast oscillations along this orbit look
very regular, hinting at a simple WKB approximation for this eigenmode.

The cardioid billiard is a stronger deformation of the circle. In this
billiard, all stable orbits have disappeared, resulting in a fully
chaotic geodesic flow (ergodic, mixing, with exponentially unstable orbits).
The center eigenmode features {\it irregular} oscillations
at the scale $\sim\lambda_n^{-1}$, but at the macroscopic scale it looks roughly
equidistributed across $\Omega$. The right eigenmode shows similar
small-scale oscillations, but it also features an {\it enhancement} of
the density $|u_n|^2$ near the vertical
periodic orbit. Such enhancements along {\it unstable} periodic orbits have
been called {\it scars} by Heller, who noticed them first
\cite{Hel84}.
Such scars along unstable orbits have still not received a
mathematical explanation, it remains unclear whether they persist, in
some form, in the limit $\lambda_n\to\infty$.

Heller's numerics were performed for the stadium billiard, which was
proved by Bunimovich to be ergodic and mixing, but which also features
a family of vertical {\it bouncing-ball} orbits, which are marginally stable.
Heller observed scars along unstable periodic orbits like for the
cardioid billiard, but also much stronger scars along the family of
vertical orbits. The persistence of those {\it strong scars} in the
high frequency limit was proven by Hassell in 2010 \cite{Hass10}. 
\begin{figure}[ht]
  \begin{center}
\includegraphics[width=0.3\tw]{
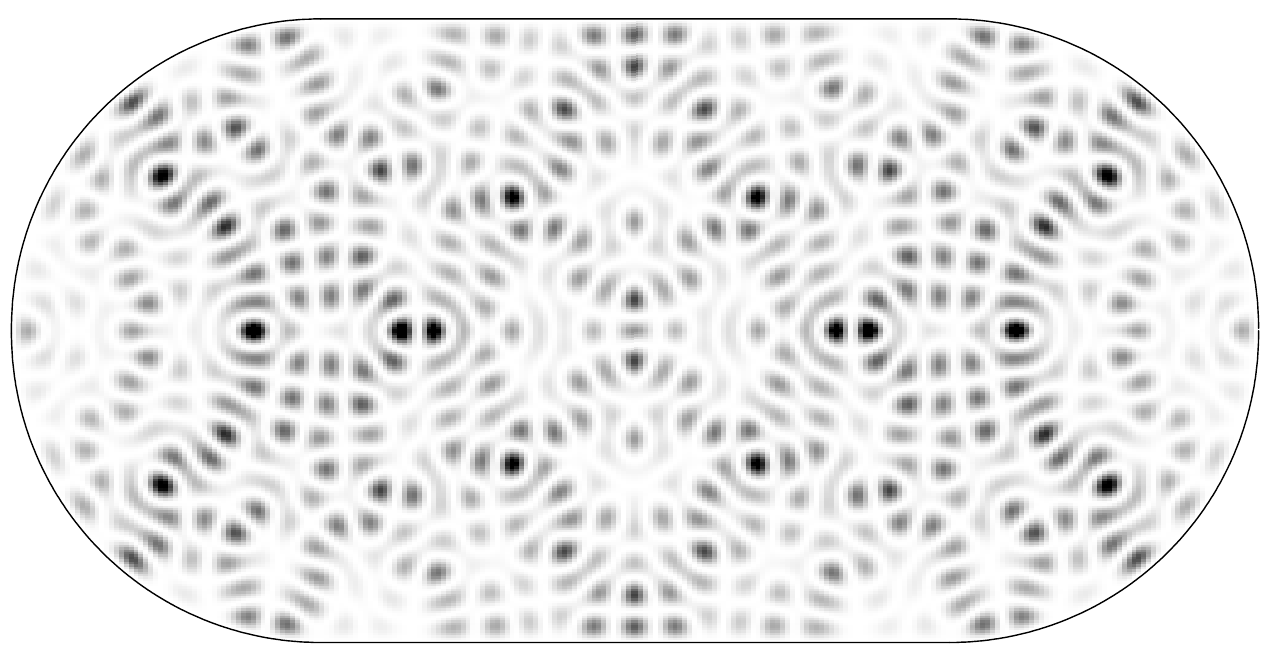} \hspace{.5cm}
\includegraphics[width=0.3\tw]{
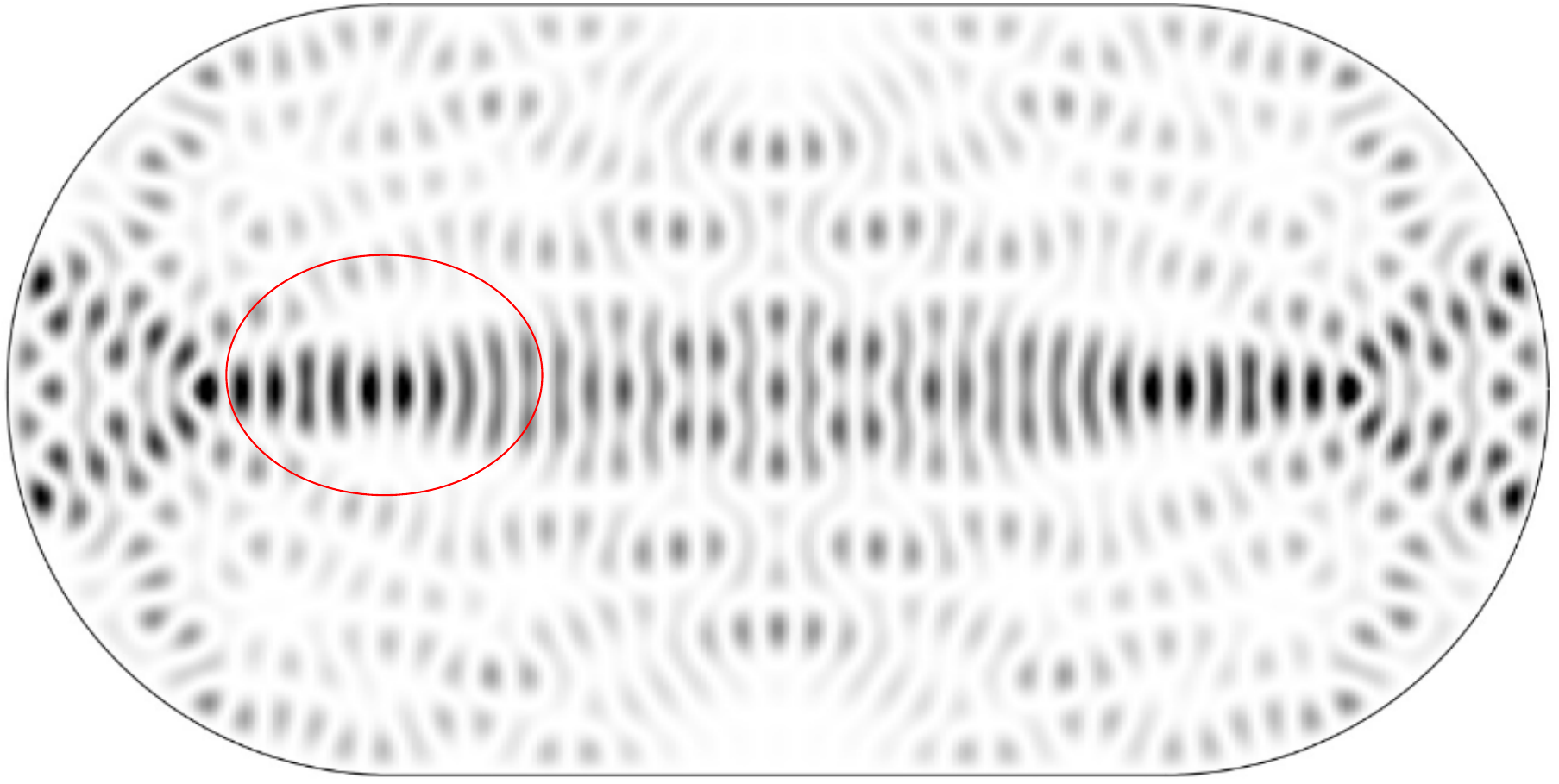} \hspace{.5cm}
\includegraphics[width=0.3\tw]{
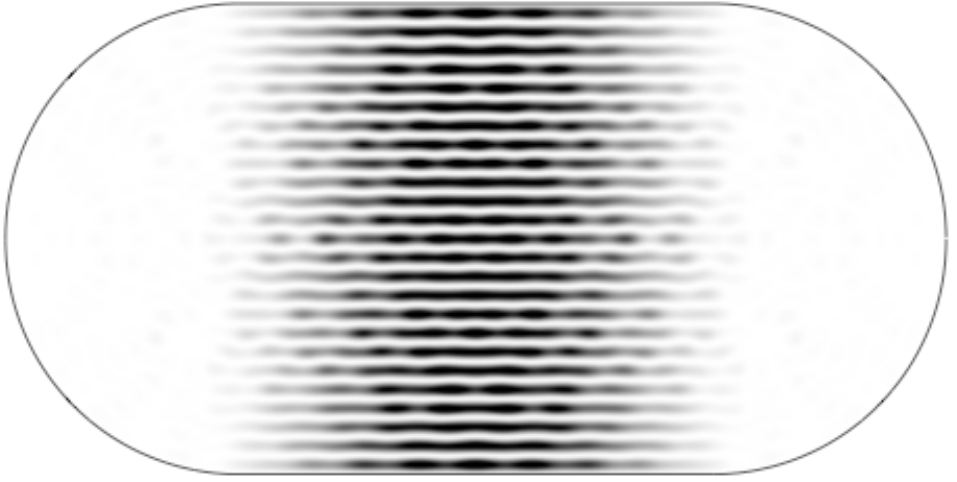}
\caption{\label{f:3modes}3 eigenmodes of the stadium
  billiard, at similar eigenfrequencies. Left: an equidistributed
  eigenmode. Center: a scar along the horizontal periodic
  orbit. Right: a strong scar along the family of bouncing ball orbits. Courtesy of Eduardo Vergini.}
\end{center}
\end{figure}

\subsection{Quantum chaos, and geometric examples}

The geodesic dynamics of the lima\c con billiard is of {\it mixed}
type: some orbits are stable, surrounded by invariant tori, while
another part of phase space appears to host a chaotic behaviour. The
separation between these two regions is in general not sharp, and
leads to complicated phenomena. The dynamics of the geodesic flow in
such mixed systems remains challenging from a mathematical point of
view.

On the opposite, the dynamics of fully chaotic billards, like the
cardioid or the stadium billards, is better understood. Most orbits
are exponentially unstable, featuring unstable and stable submanifolds
transversely to the orbit, and the foliation of phase space in to
these orbits allows to prove ergodicity and mixing of the geodesic
flow. This relatively good understanding of those chaotic billiards,
and the fact that the dynamics is homogeneous across all the phase space,
led to a special focus on the Laplacian eigenmodes of those chaotic
billiards. This is where the topic {\it Quantum Chaos} emerged, namely
the study of quantum systems whose classical counterpart is (fully)
chaotic.

At the classical level, those planar chaotic billiards are not the
simplest chaotic systems: before proving that the Sinai or stadium
billiard are chaotic, it had been known for decades that certain
geometricd models are strongly
chaotic, namely when considering
the geodesic flow on compact Riemannian surfaces $(M,g)$ of constant
negative curvature.
Indeed, on such surfaces the negative curvature is responsible for the instability of all
geodesics (hyperbolicity), and the compactness of the surface forbids the
flow from leaving to infinity, ensuring complexity. Hopf showed that
the geodesic flow on such a
surface is ergodic and mixing. In this context, the geodesic flow
$\Phi^t$ is defined on
the unit tangent bundle $SM=\{(x,v),\ x\in M, v\in
T_xM,\ ||v||=1\}$ (here $v$ is the velocity vector, tangent to $M$ at
the point $x$). Equivalently it can be defined on the unit cotangent
space $SM=\{(x,\xi),\ x\in M, \xi\in
T^*_xM,\ ||\xi||=1\}$, where $\xi$ is interpreted as the momentum,
dual to the velocity; the latter version will be more naturally
related with wave mechanics.
Ergodicity of the geodesic flow means that for almost every initial point
$(x,v)\in SM$, the orbit starting from $(x,v)$ becomes dense on $SM$,
in a uniform way. Namely, for any continuous test function $f(x,v)$ on
$SM$,
one has
\begin{equation}\label{e:ergod}
\lim_{T\to\infty}\frac{1}{T}\int_0^Tf\circ\Phi^t(x,v)\,dt = \frac{1}{\mu_L(SM)}\int_{SM}f(x,v)\,d\mu_L(x,v)\,,
\end{equation}
where $\mu_L$ is the Liouville measure on $SM$, which is the natural
volume measure on $SM$, lifting the Lebesgue measure on $M$.

In the 1960s Anosov extended those dynamical properties to
manifolds of variable negative curvature.

Compared with models of chaotic
Euclidean billiards (like the stadium billiard, or the cardioid
billiard),
the advantage of those geometric models is the
absence of boundary, making the geodesic flow $\Phi^t$ smooth
everywhere on $SM$. The uniformity of the hyperbolicity makes those
geometric Anosov flows a ``cleaner'' version of chaotic system.

At the quantum / wave level, the natural operator replacing the
Laplacian $\Delta_\Omega$ is the Laplace-Beltrami operator $\Delta_g$
associated with the metric $g$. This is the operator generating the
wave evolution on $M$, or the quantum evolution of a free particle on
$M$. This operator is naturally selfadjoint on $L^2(M,dg)$ (where $dg$
is the Lebesgue measure on $M$ associated with the metric $g$). 
Due to the compactness of $M$, this operator admits a discrete
spectrum $(u_n,\lambda_n^2)$, satisfying the analogue of Helmholtz equation:
\begin{equation}\label{e:Helm-M}
\Delta_g u_n + \lambda_n^2\,u_n=0\quad\text{on all $M$}.
\end{equation}
Eventhough this geometric model of quantum system seems less physical than that
of waves in a Euclidean box, the absence of boundary makes it easier
to analyze mathematically, and several rigorous results holding for
this geometric model are still open for waves in boxes. On the other
hand, the model of hyperbolic surfaces is very relevant certain
domains of mathematics, like number
theory and representation theory. 

At the numerical level, 
the first numerical studies of waves in chaotic cavities were devoted to 
planar billards like the stadium billiard, the Sinai billiard or the
cardioid billiard, using the fact that elementary solutions of the
wave equation are complex exponentials or Bessel functions.
Later, some groups investigated waves on hyperbolic surfaces,
e.g. D.Hejhal and his collaborators (mostly for noncompact surfaces of finite
area, like the modular surface), or F.Steiner for certain compact
hyperbolic surfaces and 3-dimensional hyperbolic manifolds, with
applications to cosmology.

\subsection{Quantum Ergodicity}
Before turning to the technical details, I will state one of the main
results obtained about Laplace eigenmodes, either on $\Omega$ (with
Dirichlet b.c.) or $M$ a compact manifold, with a simple assumption on
the geodesic flow:
\begin{theorem}[Quantum Ergodicity - spatial version]\label{thm:QE}
Let $(M,g)$ be a compact Riemannian manifold without boundary, such that the geodesic flow on $(M,g)$ is ergodic
(w.r.t. the Liouville measure on $SM$). Then there exists a
subsequence $\cS\subset\IN$ of density $1$, such that for any open
region $\omega\subset M$, one has
$$
\lim_{\cS\ni n\to\infty} \int_\omega |u_n(x)|^2\,dg(x) = \frac{\vol(\omega)}{\vol(M)}.
$$
The same statement holds if $(M,g)$ has a piecewise smooth boundary,
with an ergodic billiard flow, in particular if $(M,g)$ is an ergodic
Euclidean billiard $\Omega$.
\end{theorem}
Being ``of density $1$'' means that $\cS$ contains the vast
majority of high frequency eigenmodes:
$$
\lim_{n\to\infty}\frac{\#(\cS\cap [0,n]}{n} = 1\,.
$$
This theorem hence states that, at the macroscopic scale, ``almost
all'' the high frequency eigenmodes equidistribute across $M$. The
formal analogy with the statement of the ergodicity of $\Phi^t$ lead
to the ``Quantum Ergodicity'' nickname.

We will see later that this statement can be lifted to the phase space
$S^*M$, using pseudodifferential operators to test the phase space
distribution of the eigenstates. 

This theorem was first stated by A.Schnirelman in 1974 \cite{Schn74}, with a sketch of the
proof. Details were filled up by Zelditch (on the setup of compact
hyperbolic surfaces) \cite{Zel87} and Colin de Verdière in the more general setting
of closed Riemannian manifolds with ergodic flow \cite{CdV85}. Finally,
G\'erard-Leichtnam \cite{GL93} and Zelditch-Zworski \cite{ZZ96} dealt with Euclidean domains
and manifolds with boundaries.

The proof of Quantum Ergodicity is quite robust and could be adapted to more general settings
than that of the Laplace operator: semiclassical Schr\"odinger operators \cite{HMR87},
quantized ergodic maps \cite{BDB96}, including maps with discontinuities
\cite{MOK05,DENW06}. The notion of Quantum Ergodicity is sometimes
used also in settings where there is no underlying classical
ergodicity, e.g. to denote the asymptotic equidistribution of eigenvectors of some
ensemble of random
matrices, including random combinatorial or metric graphs
\cite{ALM15,AS19}. In the present chapter, we will stick to the
original use, namely that of quantized classical dynamical systems, in
particular geodesic flows, without any inclusion of randomness in the system.

\section{High frequency regime: phase space analysis of the eigenmodes}


In this chapter, we describe the mathematical techniques used to
analyze the high frequency regime of the Laplacian, in order to
investigate the behaviour of the eigenmodes $u_n$ when $\lambda_n\gg 1$. 
As explained earlier, the modes are oscillatory on the scale
$\lambda_n^{-1}$ (called the wavelength scale). In this regime, one is
able to connect the wave dynamics with the {\it dynamics of rays}, that is
the classical billiard dynamics inside $\Omega$, or on the manifold $(M,g)$. This connection is both singular and very
powerful, it is at the source of the main results presented below.
It is singular, because in the limit $\lambda\to\infty$, eigenmodes
become more and more oscillatory, they don't pointwise converge to any function
or distribution. The convergence we will uncover will hold only in a
weak sense, through macroscopic test functions or operators.

We will try to treat in parallel the analysis on curved manifolds
$(M,g)$ and on Euclidean billards $\Omega\subset\IR^d$. The tools used
for the analysis involve certain classes of 
{\it pseudodifferential operators}, which are naturally defined on the full
Euclidean space $\IR^d$, and need some adjustments to fit on a
manifold $M$, respectively a box $\Omega$ settings. We will only sketch those adjustments,
without giving too many details. In order to treat both systems in
parallel, we will omit the subscripts $_\Omega$ or $_g$ and only
denote the Laplacian by $\Delta$.

\subsection{High frequency vs. semiclassical regime}

To connect with classical mechanics, we use
methods from the semiclassical analysis of quantum mechanics. Indeed,
Laplacian eigenmodes also represent stationary modes of quantum massive
particles inside a Euclidean cavity (or quantum particles trapped on a
curved manifold): the Helmholtz
equation \eqref{e:Helm} is equivalent with the {\it time independent
Schr\"odinger equation}
$$
-\frac{\hbar^2 \Delta}{2m}u_n = E_n u_n\,,
$$
where $\hbar$ is Planck's constant, $m$ is the mass of the particle,
and $E_n$ its quantum eigenenergy. The large
frequency limit $\lambda_n\to\infty$ in Helmholtz equation is formally
equivalent with fixing the mass (say $m=1$), fixing the order of magnitude of the energy (e.g. taking
$E_n=E_n(\hbar)\approx 1/2$, while
sending Planck's parameter (no more ``constant'') $\hbar$ to zero:
such asymptotics is called the
semiclassical limit or, more accurately, the semiclassical regime.

This manipulation also leads us to use the semiclassical time dependent
Schr\"odinger equation
\begin{equation}\label{e:Schro}
i\hbar\frac{\partial}{\partial t}u(t,x) = H_\hbar u(t,x),\qquad  H_\hbar = -\frac{\hbar^2 \Delta}{2} u(t,x)
\end{equation}
as evolution equation, instead of the original wave
equation. Technically, the Schr\"odinger equation has the advantage to be
of order $1$ in time, and leads to a
unitary propagator acting on scalar wavefunctions $u(t)\in L^2(\Omega)$.

\subsection{Testing phase space localization}

\subsubsection{Position vs. Fourier localization}
In the statement of Thm~\ref{thm:QE}, our approach to study the eigenmodes $u_n$  consists in testing their
$L^2$ mass distribution on the physical domain $\Omega$ (or $M$), but also in Fourier
space.

Firstly, testing the mass of a (normalized) eigenmode $u_n$ inside a subdomain
$\omega\subset\Omega$ (resp. $\omega\subset M$) just amounts to computing the integral
\begin{equation}\label{e:position-dist}
\int_\omega |u_n(x)|^2\,dx = \la u_n,
\bbbone_{\omega}u_n\ra_{L^2},\quad\text{resp.}\qquad \int_\omega |u_n(x)|^2\,dg(x)
\end{equation}
More generally, one could consider smooth test functions $\chi_x\in
C^\infty(\Omega)$ instead of characteristic functions, this
smoothness will be helpful below.

In parallel, one can be interested in the mass distribution
of $u_n$ in {\it Fourier space}. This makes perfect sense in Euclidean
space, so we will first focus on this situation. One can take the standard Fourier
transform $\cF u_n$ (which is $L^2$-normalized), and test its
distribution in some region $A\subset \IR^d$, by computing $\int_A|\cF
u_n(\xi)|^2\,d\xi$. However, since the eigenmodes $u_n$ oscillate on scales $\sim
\lambda_n^{-1}$, their Fourier transform $\cF u_n$
is concentrated around Fourier parameters $|\xi|\sim
\lambda_n$. Hence, to get nontrivial informations, one needs to adapt the region $A\subset\IR^d$
to the frequency $\lambda_n$. The alternative method we will employ, consists in
rescaling the Fourier transform, to consider the {\it semiclassical Fourier
  transform}
$$
\cF_\hbar u(\xi) = (2\pi\hbar)^{-d/2}\int_\Omega
e^{-ix\cdot \xi/\hbar}\,u(x)\,d\xi,\qquad \xi\in\IR^d\,.
$$
Measuring the $L^2$ mass
of $\cF_\hbar u(\xi)$ in a fixed region $A\subset \IR^d$ amounts to computing:
\begin{align}
\int_A |\cF_\hbar u(\xi)|^2\,d\xi &= \la \cF_\hbar
                                      u,\bbbone_{A}\cF_\hbar u\ra\\
                                    &= \la u, \bbbone_{A}(\hbar D) u\ra.\label{e:momentum-dist}
\end{align}
Here we used the notation $D=-i\nabla$, so that $P=\hbar D$ is the usual
quantum momentum operator. Again, one may replace $\bbbone_A$ by a
smoother function $\chi_\xi(\xi)$ supported in $A$.
When applying the above formula to the eigenmode $u_n$, one needs to adapt the choice of $\hbar$ to the
eigenmode $u_n$ under scrutiny: to obtain a nonnegligible mass, one
needs to take $\hbar$ of the same order as $\hbar_n$, for instance
just take $\hbar=\hbar_n=\lambda_n^{-1}$. 

\subsubsection{Phase space test operators}

We see that the two expressions
(\ref{e:position-dist},\ref{e:momentum-dist}) have the same form, they
consist in bracketing a certain selfadjoint operator between the state
$u_n$. In the semiclassical limit $\hbar\ll 1$, it is possible to test
the mass distribution of eigenmodes $u_n$ {\it simultaneously} in position
and momentum space, by constructing operators from {\it phase space
  functions} $\chi\in C^\infty_c(\Omega\times \IR^d)$
depending on both
position $x\in\Omega$ and momentum\footnote{Here we
  replace the velocity $v$ with the momentum $\xi$, which is more
  central in quantum mechanics. In the Euclidean
  situation the two quantities are equivalent: $\xi=m v$. In the
  manifold setting, they are related by the metric: using Einstein's
  summation indices, $\xi^i=g^{ij}v_j$.}
$\xi\in\IR^d$. Such an operator can
be written in several ways: it can be viewed as a function of the
position and momentum operators $Q$, $P=\hbar D$, or as the
semiclassical quantization of $\chi$:
$$
\chi(Q,P)=\chi(Q,\hbar D)=\Oph(\chi).
$$
\begin{rem}
Since the operators $Q,P$ do not commute, there is a necessary choice of
{\it ordering} when defining the operator $\chi(Q,\hbar D)$. Several
choices are available, each with its own advantages. We will often
require the operator $\Oph(\chi)$ to be selfadjoint if $\chi$ is real
valued, this is the case if one uses the Weyl quantization. It is also
useful to require the positivity of the quantization: if $\chi$ is a
nonnegative function, then $\Oph(\chi)$ is a positive operator on
$L^2$.
Still, different choices of ordering (that is, of quantization) lead to differences of order $\cO(\hbar)$ between
the operators; for this reason the choice of ordering
will not be relevant for us, and we will mostly skip to indicate which
choice is made. 
\end{rem}

Operators of the type $\Oph(\chi)$ are called
$\hbar$-pseudodifferential operators. They satisfy a nice {\it
  calculus} when $\hbar\ll 1$, allowing to recover properties of the
operators $\Oph(\chi)$ from those of the functions\footnote{In
  semiclassical analysis, phase space functions are often called
  {\it symbols}.} $\chi$. One
requirement to use this calculus is that the functions $\chi$ should be smooth, so we cannot just
take $\chi=\bbbone_{B}$ for a phase space region $B$, but need to
smoothen this function a bit.

For instance, consider a phase space ball $B_0\subset \Omega\times \IR^d$ of radius
$r>0$, centered at a point $(x_0,\xi_0)$, and take $\chi_{B_0}$ a smoothened characteristic
function on that ball. Then, the bracket
\begin{equation}\label{e:bracket}
\la u_n,\chi_{B_0}(Q,\hbar_n D)u_n\ra_{L^2(\Omega)}
\end{equation}
will measure the mass of $u_n$ inside $B_0$. Notice that, as above, we have adapted Planck's parameter
$\hbar=\hbar_n$. For this mass interpretaton, it is useful that the
bracket be positive, which leads to a preference for positive
quantizations. If this bracket $\la u,\chi_{B_0}(Q,P)u\ra\approx
\|u\|^2$, this means that the particle
is essentially localized in this ball (in mathematics, one would say
{\it ``microlocalized''} to denote this simultaneous localization in
space and Fourier space).

What information do we gain from testing with
$\chi(x,\hbar D)$ instead of $\chi_x$? To give a simple example, let us consider a family of
localized plane waves
\begin{equation}\label{e:local-plane}
u_\hbar(x)=f(x)\,e^{i\xi_0\cdot x/\hbar},\quad\text{with }f\in
C^\infty_c(\Omega),\ \xi_0\in\IR^d\,.
\end{equation}
A purely spatial test function $\chi_x(x)$ will only detect the
localization properties of $f$, but will be blind to the choice of momentum
$\xi_0$. On the opposite, a phase space test function $\chi(x,\xi)$ will
detect both position and momentum localizations, as seen in the
following formulas:
$$
\la u_\hbar,\chi_x u_\hbar\ra = \int |f(x)|^2\chi_x(x)\,dx,\quad \la
u_\hbar,\Oph(\chi) u_\hbar\ra = \int |f(x)|^2\chi(x,\xi_0)\,dx +\cO(\hbar).
$$

\subsubsection{Adaptation to the cavity setting}\label{s:setting-cavity}

Unlike differential operators, pseudodifferential operators
$\chi(x,\hbar D)$ are not
local, their integral kernel is generally
supported over all of $\IR^d\times\IR^d$, so they don't preserve
$L^2(\Omega)$.
However, if the function $\chi$
is supported in the interior of $\Omega\times\IR^d$ (hence, away from
the boundary $\partial\Omega$), one can show that 
$\chi(Q,P)$ almost maps $L^2(\Omega)$ to itself, namely up to a
semiclassically very small
remainder\footnote{The notation $f(\hbar)=\cO(\hbar^\infty)$ indicates that for
any $N\geq 0$, there exists $C_N>0$ such that $f(\hbar)|\leq
C_N\hbar^N$ for any $\hbar\in (0,1]$. Hence, this quantity will be
negligible in the semiclassical limit.} $\cO(\hbar^\infty)$.

If $\supp\chi\subset \IR^d\times\IR^d$ intersects the complement of
$\Omega$, then the image $\Oph(\chi)u$ of a function $u\in
L^2(\Omega)$ will generally not be localized in $\Omega$, but will
strongly leak outside $\Omega$. One way to restore this localization property is to use a smooth
version $\psi\in C^\infty_c(\Omega)$ of the characteristic function
$\bbbone_\Omega$, and use it to cutoff the test operator as
$\psi\Oph(\chi)$ or $\psi\Oph(\chi)\psi$, see e.g. \cite{ZZ96}. With
these methods, we will not test the behaviour of the eigenmodes close to the boundary, but
rather what happens inside $\Omega$. On the opposite, \cite{GL93} used
a different approach: they ``projected'' the eigenmodes into their boundary
data, and analyzed the microlocal distribution of these boundary data in
$T^*\partial\Omega$ using test operators on the boundary. 

\subsubsection{Adaptation to the manifold setting}

The statement of Thm~\ref{thm:QE} only concerned spatial localization,
which can be defined identically on a manifold $M$ as in a Euclidean
cavity. On the other hand, the Fourier transform is not naturally defined
on a curved manifold. Except on certain specific cases (like surfaces
of constant negative curvature, for which some adapted Fourier
analysis exists), one usually applies a ``hands-on'' method by
constructing a Fourier analysis on local charts, then gluing the
resulting objects together to define a global operator $\Oph(\chi)$ on
$L^2(M)$. For a
function $\chi\in C^\infty_c(T^*M)$ (which is intrinsically defined), the operator
$\Oph(\chi)$ depends on the choice of charts and of
cutoffs used in the construction. The rules of the operator calculus
also depend on these data. However, one can show that this dependence leads to
differences of order $\cO(\hbar)$. We will therefore not specify those
choices, and keep a generic notation $\Oph(\chi)$ to denote these operators.

\subsubsection{Adjusting the Planck parameter to the eigenmode}\label{s:adapt-hn}

Let us come back to the eigenmodes $u_n$ of eigenfrequencies
$\lambda_n$. By choosing adapted Planck parameters $\hbar_n\defeq
\lambda_n^{-1}$, those eigenmodes satisfy the Helmoholtz equation
$-\hbar_n^2\Delta u_n =u_n$, which can be rewritten as
$(H_{\hbar_n}-1/2)u_n=0$. This equation implies that in
$\hbar_n$-Fourier space, the eigenmode $u_n$ is localized on the
classical energy shell $H^{-1}(1/2)=\{(x,\xi)\,:\,|\xi|=1\}$.

The need to adapt the test operator
$\chi(Q,\hbar_n D)$ to the individual eigenfrequency $\lambda_n$ may seem
unpleasant. Fortunately, this adaptation is not absolutely
necessary. Indeed, one can consider functions
$\chi(x,\xi)$ with a radial symmetry in some momentum range, e.g. in
the corona $\{1/2\leq
|\xi|\leq 2\}$, meaning that
\begin{equation}\label{e:radial}
\chi(x,r\xi)=\chi(x,\xi),\qquad \text{whenever}\ |\xi|=1\
\text{and}\quad r\in [1/2,2].
\end{equation}
For such functions $\chi$, as long as the ratio
$\hbar/\hbar_n$ remains in the interval $[1/2,2]$, one can show that
\begin{equation}\label{e:h-hn}
\chi(Q,\hbar D)u_n = \chi(Q,\hbar_n D)u_n+\cO(\hbar^\infty).
\end{equation}
So for such a ``locally radial'' function $\chi$, one may use the same parameter $\hbar$ to test
the localization of all eigenmodes in the frequency range
$\lambda_n\in [1/2\hbar^{-1},2\hbar^{-1}]$.

Functions of the radial type \eqref{e:radial} can be constructed from
functions defined over the unit cotangent
bundle  $\tchi\in C^\infty(S^*M)$, then extended radially throughout a corona
$\{|\xi|\in [1/4,4]$ and cutoff by some radial function
$\psi\in C^\infty(\IR)$ vanishing outside $[1/4,4]$ and equal to unity in
$[1/2,2]$:
\begin{equation}\label{e:radial-ext}
\chi(x,\xi) \defeq \psi(|\xi|)\,\tchi(x,\xi/|\xi|),\forall (x,\xi)\in T^*M\,.
\end{equation}.

\subsection{Measuring the $L^2$ mass of eigenmodes in mesoscopic phase
  space balls (KEEP FOR LATER?)}

Above we have taken a test function $\chi\in
C^\infty_c(\Omega\times\IR^d)$ supported in a ball
$B_0\subset\Omega\times\IR^d$, chosen
independently of the semiclassical parameter $\hbar$. Such
$\hbar$-independent test functions measure the {\it macroscopic} mass distribution
of $u_n$, they are unable to test the oscillations of $u_n$ at the {\it
  microscopic} scale (or {\it Planck scale}) $\hbar_n=\lambda_n^{-1}$
we have alluded to earlier. In some sense, this information is very
{\it rough}, it is unable to describe the fine structure of the
eigenmodes $u_n$. 

Could one use test functions
$\chi=\chi_\hbar$ supported in balls with decaying radii
$r(\hbar)\searrow 0$? Operators $\Oph(\chi_\hbar)$ still makes sense,
and benefit from the $\hbar$-pseudodifferential
calculus, provided the functions oscillates on scales $\gg
\hbar^{1/2}$. So one could consider testing the mass distribution
inside {\it mesoscopic} balls $B_\hbar$ of radii $r_\hbar\gg\hbar^{1/2}$.

\subsection{Quantum-classical correspondence: mass propagation on $(M,g)$}

So far we have described how to test the phase space mass
distribution of our eigenmodes. Let us now show how to make use of the
fact that the $u_n$ are eigenmodes of $\Delta$, by relating
classical and quantum (or wave) flows. For a moment we will consider
only the setting of a manifold without boundary.

We have seen that the Laplacian $\Delta$ can be used to generate the
semiclassical Schr\"odinger equation \eqref{e:Schro} (setting $m=1$). 
The solution to this equation is provided by the unitary group
(unitary {\it propagator})
$U^t_\hbar= \exp(-itH_\hbar/\hbar)$, $t\in\IR$. Hence, for a given
wavefunction $u_0\in L^2$, the evolved wavefunction at time $t$ is given
by $u_t=U^t_\hbar u_0$. If we use a test operator $\Oph(\chi)$ to test the
distribution of wavefunctions in phase space, how can we relate the
mass distribution of $u_0$ and that of $u_t$?

On the classical side, the geodesic flow is equivalent with the
Hamiltonian flow $\Phi^t$ generated by the
classical Hamiltonian
$$
H(x,\xi)=\frac{|\xi|^2}{2},
$$
where
$|\xi|^2=g(\xi,\xi)$ is the norm in the cotangent space generated by
the metric $g$. This flow is homogeneous: the trajectories of the
points $(x_0,\xi_0)$ and $(x_0,r\xi_0)$ are related to one another by
multiplying the momenta, and the speed of evolution, by the factor
$r$. 
In order to study the dynamical properties of the flow, it is
therefore sufficient to consider the energy
shell $H^{-1}(1/2)$, which amounts to the unit cotangent bundle
$S^*M=\{(x,\xi),\ |\xi|=1\}$.

The correspondence between the quantum and classical evolutions takes the
form of the following estimate:
\begin{equation}\label{e:Egorov}
U_\hbar^{-t}\,\Oph(\chi)\, U_\hbar^t = \Oph(\chi\circ\Phi^t)+\cO_t(\hbar)
\end{equation}
In mathematics this type of equality is referred to as Egorov's
theorem. The remainder $\cO_t(\hbar)$ estimates the operator norm of
the difference between the two sides. The suffix $_t$ indicates that
the implied constant depends on the time (typically, in an exponential
way).

If we sandwich this operator between some initial wavefunction $u_0$,
we see that it takes the form:
$$
\la u_t,\Oph(\chi)\, u_t\ra= \la u_0, \Oph(\chi\circ\Phi^t) u_0\ra+\cO_t(\hbar)\,.
$$
If $\chi\approx \bbbone_B$, then the LHS measures the mass of $u_t$
inside a phase space ball $B$. Then $\chi\circ\Phi^t\approx
\bbbone_{\Phi^{-t}(B)}$, meaning that the RHS measures the mass of
$u_0$ inside the backwards evolved ball $\Phi^{-t}(B)$. The above
estimate shows that these two masses are approximately equal to each
other. This is intuitive: if $u_0$ is concentrated microlocally inside
the set $B_{-t}=\Phi^{-t}(B)$, then $u_t$ is concentrated in
$\Phi^t(B_t)=B$. In this sense, the mass distribution of wavefunctions
evolves like the mass distribution of a cloud of classical particles.

This quantum-classical correspondence allows to relate the mass of
eigenmodes $u_n$ along different points of a geodesic. Indeed, keeping
the same notations and assumptions as above, the fact that $u_n$ is an eigenmode of
the propagator $U^t_\hbar$ (here we take $\hbar=\hbar_n$) implies
that:
\begin{equation}\label{e:mass-propag}
\la u_n,\Oph(\chi)u_n\ra = \la U_\hbar^tu_n,\Oph(\chi)U_\hbar^tu_n\ra
= \la u_n,\Oph(\chi\circ\Phi^t)u_n\ra + \cO(h)\,.
\end{equation}
If $\chi\approx \bbbone_{B_0}$ then $\chi\circ\Phi^t\approx
\bbbone_{\Phi^{-t}(B_0)}$. This shows that the $L^2$ mass of $u_n$ in
a phase space ball $B_0$ approximately equals the mass of $u_n$ in
$\Phi^{-t}(B_0)$. This property is sometimes called ``mass
propagation'' along the geodesic flow.

This ``mass propagation'' mimicks the properties of a $(\Phi^t)$-invariant measure
$\mu$ on $T^*M$: the flow-invariance precisely means that for any
measurable set $B\subset T^*M$ and any $t\in\IR$, $\mu(\Phi^{-t}(B))=\mu(B)$. 

This leads us to the important notion of semiclassical measures
associated with the sequence of eigenmodes $(u_n)$. 

\section{Semiclassical measures for the Laplacian on $(M,g)$}

There is indeed a way to associate certain $\Phi^t$-invariant measures
to the family of eigenmodes $(u_n)_{n\geq 1}$ of the Laplacian on
$M$. As we have seen above, for a test
function $\chi\in C^\infty_c(T^*M)$ approximating the characteristic
function of a region $B\subset T^*M$, the $L^2$ bracket
$\la u_n,\Op_{\hbar_n}(\chi) u_n\ra$ approximately measures the mass of
$u_n$ in $B$. Because the quantization $\chi\mapsto \Oph(\chi)$ is a
linear mapping, the map
$$
\chi\in C^\infty_c(T^*M)\mapsto \la u_n,\Op_{\hbar_n}(\chi) u_n\ra
$$
is also linear, so it can be interpreted as a distribution $\mu_n$ on
$T^*M$. Let us assume that we have chosen the quantization procedure
to be positive. The distribution $\mu_n$ is then itself positive,
namely it is a positive Borel measure on $T^*M$.
This measure actually extends to all
The quantization maps bounded symbols $\chi\in
C^\infty_b(T^*M)$ to bounded operators, so the measure $\mu_n$ extends
to bounded functions $\chi$, in particular the constant function
$\chi\equiv 1$.
The quantization procedure satisfies $\Oph(1)=Id$, which shows that $\mu_n$ is
a probability measure on $T^*M$.

The Helmholtz equation $(H_{\hbar_n}-1/2)u_n = 0$, together with
the fact that $H_\hbar = \Oph(H) (1+\cO(h))$, with $H(x,\xi)=|\xi|^2/2$, shows that the
measures $\mu_n$ are asymptotically concentrated on the energy shell
$H^{-1}(1/2)=S^*M$. Indeed, if $\chi\in C^\infty_b(T^*M)$ vanishes
near $H^{-1}(1/2)$, the function
$g=\frac{\chi}{H-1/2}$ is smooth and bounded. The pseudodifferential calculus then shows
that
$$
\Oph(\chi) =\Oph(g) (H_\hbar-1/2) +\cO(\hbar) \,,
$$
When we apply this expression (with $\hbar=\hbar_n$) to the eigenmode $u_n$, we get
$\mu_n(\chi)=\cO(\hbar_n)$. This shows that the mass of
$\mu_n$ outside $S^*M$ is asymptotically vanishing when $n\to\infty$. 

Depending on the choice of quantization we make, the measures $\mu_n$ are
called Wigner distributions (which are not positive in general), or
Husimi measures (which are positive, hence measures). If the
mass propagation estimate \eqref{e:mass-propag} was exact, the measure
$\mu_n$ would be invariant through the geodesic flow $\Phi^t$. But
because of the $\cO(\hbar)$ remainder, it is only approximately
invariant.

In order to obtain true invariant measures, the idea is to take limits of the measures
$(\mu_n)_{n\geq 1}$. These limits are to be understood in a weak-$*$
sense, meaning that for any test function $\chi$, the expressions
$\mu(\chi)\in\IR$ have a limit in $\IR$.

A priori, successive measures $\mu_n$, $\mu_{n+1}$, $\mu_{n+2}$ do
resemble each other. So well-defined limits will exist only if one
extracts adequate subsequences of eigenmodes. One can show that,
indeed, one can always extract a subsequence $(\mu_{n_k})_{k\geq 1}$
such that the measures $(\mu_{n_k})$ have a limit when $k\to\infty$, in
this weak-$*$ sense:
$$
\forall \chi\in C^\infty_b(T^*M),\quad \la
u_n,\Op_{\hbar_n}(\chi)u_n\ra \stackrel{k\to\infty}{\to} \mu_{sc}(\chi)\,.
$$
Here $\mu_{sc}$ is the weak-$*$ limit of the $(\mu_{n_k})$, it
reflects the asymptotic macroscopic distribution of the measures $\mu_n$ across $T^*M$. It is a probability measure on $T^*M$, called the {\it
  semiclassical measure} associated with the subsequence $(u_{n_k})$.
This measure $\mu_{sc}$ is exactly supported on $S^*M$, and it is
exactly invariant through the geodesic flow $\Phi^t$. 

Semiclassical measures are therefore limit points of the sequence of measures
$(\mu_n)_{n\geq 1}$. Identifying all the semiclassical measures 
amounts to understanding all possible phase space distributions of the
eigenmodes $u_n$, in the high frequency limit. In particular, since
semiclassical measures are necessarily flow-invariant, one may wonder
whether any flow-invariant measure $\mu$ can be obtained as a
semiclassical measure. In short: can high frequency eigenmodes
distribute according to any given invariant measure $\mu$, or does
wave mechanics select a strict subset of invariant measures?

\subsection{Quantum Ergodicity in phase space}
On a general manifold, this question is difficult. For a manifold $M$
hosting an ergodic geodesic flow, Quantum Ergodicity, when lifted to
phase space, provides a
partial answer to this question:
\begin{theorem}[Quantum Ergodicity - phase space version]\label{thm:QE2}
Let $(M,g)$ be a compact Riemannian manifold without boundary, such that the geodesic flow on $(M,g)$ is ergodic
(w.r.t. the Liouville measure on $S^*M$). Then there exists a
subsequence $\cS\subset\IN$ of density $1$, such that for any $\chi\in
C^\infty_b(T^*M)$, 
$$
\lim_{\cS\ni n\to\infty} \la u_n,\Op_{\hbar_n}(\chi)u_n\ra =
\int_{S^*M} \chi\,d\mu_{L},
$$
where $\mu_L$ is the Liouville probability measure on $S^*M$.

The same statement holds if $(M,g)$ has a piecewise smooth boundary,
with an ergodic billiard flow, in particular if $(M,g)$ is an ergodic
Euclidean billiard.
\end{theorem}
The spatial version of Quantum Ergodicity theorem~\ref{thm:QE2} is a
consequence of the one above. Indeed, taking test functions $\chi\in
C^\infty_b(T^*M)$ independent of the momentum
variable, the operator $\Oph(\chi)$ is equal to the multiplication by
$\chi(x)$, which leads to
$$
\lim_{\cS\ni n\to\infty} \int_M \chi(x)\,|u_n(x)|^2\,dg(x) =
\int_{S^*M} \chi\,d\mu_{L} = \frac{1}{\vol(M)}\int_M \chi(x)\,dg(x)\,.
$$
In the language of semiclassical measures, the theorem says that the
subsequence $(\mu_{n})_{n\in\cS}$ admits a single limit point, which
is the Liouville measure. So, the Liouville measure is a semiclassical
measure, associated with a ``very large'' subsequence of eigenmodes.

\subsection{Quantum Ergodicity: elements of proofs}

Depending on the specific system under study, the proofs of the
quantum ergodicity theorem have used slightly different methods, or
test operators. Since we have used above a semiclassical formalism, we
will keep to this formalism, borrowing several steps from \cite{HMR87}, instead of using the ``homogeneous
pseudodifferential calculus'' as in \cite{CdV85} or \cite{Zel87}. 

\subsubsection{Semiclassical  Weyl's law}
To start, one needs some more informations on the $\hbar$-pseudodifferential
operators, namely some properties of their traces. Taking a test
function $\chi\in C^\infty_c(T^*M)$, one can show that the operator
$\Oph(\chi)$ is a trace class operator on $L^2(M)$, and in the limit $\hbar\ll 1$ its trace
is related with the phase space integral of the symbol $\chi$:
\begin{equation}\label{e:trace}
\tr(\Oph(\chi)) = \frac{1}{(2\pi\hbar)^d}\int_{T^*M}
\chi(x,\xi)\,dx\,d\xi+ \cO(\hbar^{-d+1})\,.
\end{equation}
Notice that the integration measure $dx\,d\xi$ can be defined in each
coordinate chart of
$M$, but the measure is actually independent of the chart (it is the
natural symplectic measure on $T^*M$).

This asymptotic formula is important, it relates traces (which can be
viewed as ``quantum sums'') with classical integrals over phase
space (``classical sums'').

Another important ingredient of the $\hbar$-calculus is the so-called
functional calculus. Namely, if we recall that $H_\hbar$ is the
quantization of $H(x,\xi)=\|\xi\|^2/2$, and consider a smooth real
valued bounded
function $f\in C^\infty_b(\IR)$, then the operator $f(H_\hbar)$
(which can be defined through spectral theory) is close to the
pseudodifferential operator $\Oph(f\circ H)$:
\begin{equation}\label{e:functional}
f(H_\hbar)=\Oph(f\circ H)+\cO(\hbar).
\end{equation}
The combination of this functional calculus and of the trace estimate \eqref{e:trace} allows to approximately count the number of eigenvalues
of $H_\hbar$ in a give interval $[\alpha,\beta]$, where we choose
$0<\alpha\leq 1/2< \beta$ as $\hbar$-independent values.
Namely, if one
defines two functions $f_{\pm}$ satisfying:
$$
\bbbone_{[\alpha+\eps,\beta-\eps]}\leq f_\leq \leq \bbbone_{[\alpha,\beta]}\leq
f_+(t)\leq \bbbone_{[\alpha-\eps,\beta+\eps]},
$$
then we may compare the selfadjoint operators
$$
\bbbone_{[\alpha+\eps,\beta-\eps]}(H_\hbar)\leq f_-(H_\hbar)\leq
\bbbone_{[\alpha,\beta]}(H_\hbar)\leq f_+(H_\hbar)\leq \bbbone_{[\alpha-\eps,\beta+\eps]}(H_\hbar),
$$
and hence also their corresponding traces.
$$
\tr(f_-(H_\hbar))\leq \tr(\bbbone_{[\alpha,\beta]}(H_\hbar))\leq \tr(f_+(H_\hbar)).
$$
We then notice that, for any interval $I$, the trace $\tr \bbbone_{I}(H_\hbar)$ exactly counts the eigenvalues
of $H_\hbar$ in the interval $I$:
$$
\tr(\bbbone_{I}(H_\hbar)) = \#\{\spec(H_\hbar)\cap I\}.
$$
On the other hand, the functional calculus gives estimates on the 
traces of $f_{\pm}(H_\hbar)$:
$$
\tr(f_{-}(H_\hbar)) = \frac{1}{(2\pi\hbar)^d}\int_{T^*M}
f_-(H(x,\xi))\,dx\,d\xi+ \cO(\hbar^{-d+1})\geq \frac{1}{(2\pi\hbar)^d}\int_{T^*M}
\bbbone_{[\alpha+\eps,\beta-\eps]}(H(x,\xi))\,dx\,d\xi+ \cO(\hbar^{-d+1}),
$$
and similarly for $\tr(f_{+}(H_\hbar))$. The integral on the RHS gives
us the phase space volume of the energy layer $H^{-1}([\alpha+\eps,\beta-\eps])$.
We finally get the following bounds when $\hbar\to 0$:
\begin{equation}\label{e:2bounds}
\frac{1}{(2\pi\hbar)^d}\vol_{T^*M}\Big(H^{-1}([\alpha+\eps,\beta-\eps])\Big)+
\cO(\hbar^{-d+1})
\leq \#\{\spec(H_\hbar)\cap [\alpha,\beta]\}
\leq
\frac{1}{(2\pi\hbar)^d}\vol_{T^*M}\Big(H^{-1}([\alpha-\eps,\beta+\eps])\Big)+
\cO(\hbar^{-d+1})\,.
\end{equation}
Due to the fact that $H(x,\xi)=|\xi|^2/2$, the volumes of the phase
space region $H^{-1}([\alpha,\beta])$ take the simple value
$$
\vol_{T^*M}\Big(H^{-1}([\alpha,\beta])\Big) =  ((2\beta)^{d/2}-(2\alpha)^{d/2})\Omega_d\vol(M),
$$
where $\Omega_d=\frac{2\pi^{d/2}}{d\Gamma(d/2)}$ is the volume of the
unit ball in $\IR^d$. This shows that for $\eps>0$ small, the two
above integrals can be estimated as
$$
\vol_{T^*M}\Big(H^{-1}([\alpha\mp\eps,\beta\pm\eps])\Big) = ((2\beta)^{d/2}-(2\alpha)^{d/2})\Omega_d\vol(M) + \cO(\eps)
$$
We may actually let $\eps$
depend on $\hbar$, say like some power $\eps=\hbar^\delta$ for some
exponent $\delta>0$. This
forces to take more singular, $\hbar$-dependent cutoff functions
$f_{\pm}$, with derivative estimates
$\|f_{\pm}\|_{C^m}=\cO(\hbar^{-m\delta})$.  
The functional calculus can still be applied to such singular
functions, as long as $\delta<1/2$. One then obtains, for such a fixed interval $[\alpha,\beta]$:
$$
\#\{\spec(H_\hbar)\cap [\alpha,\beta]\} = \frac{1}{(2\pi\hbar)^d} ((2\beta)^{d/2}-(2\alpha)^{d/2})\Omega_d\vol(M) + \cO(\hbar^{-d+\delta})
$$
By using more sophisticated methods (using smoothened characteristic functions $f$
varying on scales $~\hbar$), one is able to obtain
a more precise remainder:
\begin{equation}\label{e:Weyl-opt}
\#\{\spec(H_\hbar)\cap [\alpha,\beta]\} =  \frac{1}{(2\pi\hbar)^d} ((2\beta)^{d/2}-(2\alpha)^{d/2})\Omega_d\vol(M) +
\cO(\hbar^{-d+1})\,.
\end{equation}
This estimate can be carried out even if the interval
$I=[\alpha,\beta]$ itself
depends on $\hbar$, for instance by taking intervals:
$$
[\alpha(\hbar),\beta(\hbar)]=[\frac{1}{2}, \frac{1}{2}+w(\hbar)]\,,
\quad\text{with a width}\ w(\hbar)=\hbar^{\delta},\ \ 0\leq \delta\,.
$$
The trace estimate remains informative as long as the main term in
\eqref{e:Weyl-opt} dominates
the remainder, that is if $1\gg w(\hbar)\gg\hbar$. In such short intervals, the
counting estimate \eqref{e:Weyl-opt} still predicts that the number of
eigenvalues in the intervals grows to infinity when $\hbar\to 0$:
$$
\#\{\spec(H_\hbar)\cap [\frac{1}{2}, \frac{1}{2}+w(\hbar)]\} =
\frac{w(\hbar)}{(2\pi\hbar)^d}\omega_{d-1}\vol(M) + \cO(\hbar^{-d+1})\,,
$$
where $\omega_{d-1}=d\Omega_d$ is the volume of the
$(d-1)$-dimensional unit sphere.

These semiclassical estimates were obtained for more general
Hamiltonian operators $H_\hbar$ on $\IR^d$ \cite{HMR87}. For the
Laplacian on a Riemannian manifold, they amount to the original
counting asymptotics first obtained by Weyl on Euclidean domains \cite{Wey11}, then
improved and adapted to Riemannian manifolds by Levitan
\cite{Lev52,Lev55}, Avakumovic \cite{Ava56} and
H\"ormander \cite{Hor68} by using various methods. In the case of closed
Riemannian
manifolds without conjugate points (e.g. manifolds of negative
curvature), B\'erard \cite{Ber77} showed that the remainder in \eqref{e:Weyl-opt}
can be improved to $\cO(\hbar^{-d+1}|\log \hbar|^{-1})$.

\begin{rem}
  Since $H_\hbar=-\hbar^2\Delta/2$, the condition
  $E_{\hbar,j}\in [\alpha,\beta]$ amounts to the fact the index $j$
  belongs to the set
\begin{equation}\label{e:Jh}
  J_\hbar=J_\hbar([\alpha,\beta])\defeq \{j\in\IN,\ \lambda_j\in
[\hbar^{-1}\sqrt{2\alpha},\hbar^{-1}\sqrt{2\beta}]\}.
\end{equation}
\end{rem}

\subsubsection{Generalized Weyl's law}

The trace estimates allow us not only to count eigenvalues in an
interval $[\alpha,\beta]$, but also to study the phase space
distribution of the corresponding eigenmodes $u_j$.

Indeed, we can multiply the energy cutoffs $f_{\pm}(H_\hbar)$ used to count the
eigenvalues by a pseudodifferential
$\Oph(\chi)$ obtained by quantizing an arbitrary test function $\chi\in
C^\infty_c(T^*M)$. The resulting operator is still a
pseudodifferential operator, with principal symbol $\chi\times f\circ
H$.
This trace can be expanded into
$$
\tr \Oph(\chi)f(H_\hbar) =\sum_{j} f(E_{\hbar,j})\,\la u_j,\Oph(\chi)u_j\ra\,,
$$
where the eigenenergies of $H_\hbar$ are simply given by $E_{\hbar,j}=\hbar^2\lambda_j/2$.
In quantum mechanics, each bracket (or ``matrix element'') $\la u_j,\Oph(\chi)u_j\ra$ is
interpreted as the ``quantum average'' of the operator $\Oph(\chi)$
w.r.t. the eigenstate $u_j$.

Because we use the same operator $\Oph(\chi)$ for all states $u_j$ in the
spectral range $\{E_{\hbar,j}\in \supp f\}$, those eigenstates will
not all be microlocalized in the same energy shell: eigenstates with
$E_{\hbar,j}\approx E$ will be microlocalized in the shell
$H^{-1}(E)$. 
The role of $\Oph$ is to determine how the $u_j$ are distributed
across within the shell $H^{-1}(E)$. It is then convenient to use
``locally radially symmetric'' functions $\chi$ as described in
section~\ref{s:adapt-hn}, which have the same dependence w.r.t. $x$
and the momentum direction $\xi/|\xi|$ within the layer $|\xi|\in
[1/2,2]$, equivalently within the energy layer $E\in [1/8,2]$. 

If the spectral intervals
$[\alpha,\beta]\subset [1/8,2]$, then the equality \eqref{e:h-hn} shows
that for any eigenstate $u_j$ with eigenvalue $E_{\hbar,j}\in
[\alpha,\beta]$, the brackets
\begin{equation}\label{e:h-hn2}
\la u_j,\Oph(\chi)u_j\ra = \la u_j,\Op_{\hbar_j}(\chi)u_j\ra + \cO(\hbar^\infty)\,.
\end{equation}
This just indicates that measuring the microlocalization of $u_j$
within the shell $S^*M=H^{-1}(1/2)$, using the adapted operator
$\Op_{\hbar_j}(\chi)$, is equivalent with measuring its
microlocalization 
within the shell $H^{-1}(E_{\hbar,j})$ using the $\Oph(\chi)$.

We are now ready to apply the trace asymptotics \eqref{e:trace} to the
operator $\Oph(\chi)f(H_\hbar)=\Oph(\chi\, f\circ H)+\cO(\hbar)$:
\begin{equation}\label{e:GWL1}
\tr \Oph(\chi)f(H_\hbar) = \frac{1}{(2\pi\hbar)^d}\int_{T^*M}
\chi(x,\xi) f\circ H(x,\xi)\,dx\,d\xi + \cO(\hbar^{-d+1})\,.
\end{equation}
Thanks to the local radial property of $\chi$, the integral on the RHS
decouples between a radial and angular momentum parts:
$$
\int_{T^*M}\chi(x,\xi) f\circ H(x,\xi)\,dx\,d\xi = \int_0^\infty f(r^2/2)\,r^{d-1}dr\int_{S^*M}\chi(x,\omega)\,dx\,d\omega\,,
$$
where $\omega=\xi/|\xi|$ is the momentum angular variable. The last
integral involves a multiple of the normalized Liouville measure
$\mu_L$ on $S^*M$:
$$
\int_{S^*M}\chi(x,\omega)\,dx\,d\omega = \vol(M)\omega_{d-1}\int_{S^*M}\chi\,d\mu_L\,,
$$
where $\omega_{d-1}=d\Omega_d$ is the volume of the unit sphere
$S^{d-1}$.

Playing with functions $f_{\pm}$ approximating $\bbbone_{[\alpha,\beta]}$ on
scales $\hbar$ as explained above, and using \eqref{e:h-hn2}, one can show the
following {\it generalized Weyl's law}:
\begin{equation}\label{e:generalizedWL}
  \sum_{j\in J_\hbar([\alpha,\beta])}
  \la u_j,\Op_{\hbar}(\chi)u_j\ra = \frac{1}{(2\pi\hbar)^d}
((2\beta)^{d/2}-(2\alpha)^{d/2})\Omega_d\vol(M)
\int_{S^*M}\chi\,d\mu_L + \cO(\hbar^{-d+1})\,.
\end{equation}
Comparing this expression with the Weyl counting formula
\eqref{e:Weyl-opt}, the above expression can be viewed as the {\it
  average}
of the matrix elements $( \la u_j,\Op_{\hbar}(\chi)u_j\ra )_{j\in J_\hbar( [\alpha,\beta])}$. Taking a probabilistic
viewpoint, we write this average
\begin{equation}\label{e:average}
\IE_{J_\hbar}\big[ \la u_j,\Op_{\hbar}(\chi)u_j\ra \big] = \int_{S^*M}\chi\,d\mu_L+\cO(\hbar)\,.
\end{equation}
Hence the average of the quantum matrix elements converges to the
classical average of the function $\chi$ over phase space. We will
denote this classical average $\mu_L(\chi)=\int_{S^*M}\chi\,d\mu_L$.

\subsubsection{The quantum variance}\label{s:variance}
Now that we computed the average of the distribution of matrix
elements, the next goal is to estimate the {\it variance} of this
distribution. More precisely, we will try to compute the second moment
of the distribution of elements $(\la u_j,\Op_{\hbar}(\chi)u_j\ra -
\int_{S^*M}\chi\,d\mu_L)_{j\in J_\hbar}$. We will denote by  the classical average of $\chi$.

In order to compute this variance, we will use the invariance of these
elements through the quantum propagator $U_\hbar^t=e^{-iH_\hbar t/\hbar}$: due to the fact
that $u_j$ are eigenstates of $H_\hbar$, we have
\begin{align}\label{e:qu-ave}
\forall t\in \IR,\qquad \la u_j,\Oph(\chi)u_j\ra &= \la U_\hbar^tu_j,\Oph(\chi)U_\hbar^tu_j\ra \\
\Longrightarrow  \la u_j,\Oph(\chi)u_j\ra &= \frac{1}{T}\int_0^T\la
                                            U_\hbar^tu_j,\Oph(\chi)U_\hbar^tu_j\ra\,dt\\
                                                 &\defeq\la u_j,\overline{\Oph(\chi)}^T u_j\ra,
\end{align}
where we defined the quantum average of $\Oph(\chi)$ as:
\begin{equation}\label{e:qu-average}
\overline{\Oph(\chi)}^T=\frac{1}{T}\int_0^TU_\hbar^{-t}\Oph(\chi)U_\hbar^t\,dt\,.
\end{equation}
Using the quantum classical correspondence \eqref{e:Egorov}, we end up with
$$
\la u_j,\Oph(\chi)u_j\ra =\la u_j, \overline{\Oph(\chi)}^T u_j\ra + 
\cO_T(\hbar) =  \la u_j,\Oph(\overline{\chi}^T) u_j\ra + 
\cO_T(\hbar)\,,
$$
where $\overline{\chi}^T$ is the time average of the function $\chi$ under the
geodesic flow up on the time interval $[0,T]$.

After these manipulations, let us estimate the following second
moment over the spectral interval $[\alpha,\beta]$:
\begin{align*}
 \sum_{j\in J_\hbar}\Big| \la u_j,\Oph(\chi)u_j\ra - \mu_L(\chi) \Big|^2 &=
 \sum_{j}\Big| \la u_j,\Oph(\overline{\chi}^T - \mu_L(\chi)
 )u_j\ra + \cO_T(\hbar) \Big|^2 \\
 &\leq \sum_{j\in J_\hbar}\Big| \la u_j,\Oph(\overline{\chi}^T - \mu_L(\chi)
 )u_j\ra \Big|^2 + \cO_T(\hbar) \\
 &\leq \sum_{j\in J_\hbar}\Big\|\Oph(\overline{\chi}^T - \mu_L(\chi)
 )u_j \Big\|^2+ \cO_T(\hbar)\\
 &= \sum_{j\in J_\hbar}\la u_j, \Oph(\overline{\chi}^T - \mu_L(\chi)
 )^*(\Oph(\overline{\chi}^T - \mu_L(\chi)
 )u_j \ra + \cO_T(\hbar)\,,
\end{align*}
where we just used a Cauchy-Schwarz inequality in the third line. Note
that the remainder $\cO_T(\hbar)$ is present in each $j$ term.

Now, the pseudodifferential calculus allows to replace the product of
operators by
$$
\Oph(\overline{\chi}^T - \mu_L(\chi))^*(\Oph(\overline{\chi}^T -
\mu_L(\chi)) = \Oph(|\overline{\chi}^T - \mu_L(\chi)|^2)+\cO_T(\hbar)\,.
$$
so we get:
$$
\sum_{j\in J_\hbar}\Big| \la u_j,\Oph(\chi)u_j\ra - \mu_L(\chi)
\Big|^2 \leq 
\sum_{j\in J_\hbar} \la u_j, \Oph(|\overline{\chi}^T - \mu_L(\chi)|^2)u_j\ra +\cO_T(\hbar)
$$
Applying the generalized Weyl's formula to this expression, we finally
get:
\begin{equation}\label{e:GWL-T}
\sum_{j\in J_\hbar}\Big| \la u_j,\Oph(\chi)u_j\ra - \mu_L(\chi)
\Big|^2 \leq 
\frac{1}{(2\pi\hbar)^d}
((2\beta)^{d/2}-(2\alpha)^{d/2})\Omega_d\vol(M)
\int_{S^*M}|\overline{\chi}^T - \mu_L(\chi)|^2\,d\mu_L + \cO_T(\hbar^{-d+1})\,.
\end{equation}
Dividing this expression by the counting
$\# J_\hbar$, we obtain the following
expression relating the {\it
  quantum variance} (the variance of the brackets $\la
u_j,\Oph(\chi)u_j\ra$) to the classical variance for the value
distribution of the time averaged
function $\overline{\chi}^T$:
\begin{equation}\label{e:qu-variance}
  \Var_\hbar([\alpha,\beta])\defeq \frac{1}{\# J_\hbar([\alpha,\beta])
  }
  \sum_{j\in J_\hbar}\Big| \la u_j,\Oph(\chi)u_j\ra - \mu_L(\chi)
  \Big|^2 = \int_{S^*M}\big|\overline{\chi}^T - \mu_L(\chi)\big|^2\,d\mu_L + \cO_T(\hbar)\,.
\end{equation}

\subsubsection{Making use of the ergodicity assumption}

So far we have not made any assumption on the geometry of $(M,g)$, and
the dynamical properties of the geodesic flow $\Phi^t$.
Let us now make the assumption that the flow is {\it ergodic} on $S^*M$,
with respect to the invariant probability measure $\mu_L$. In
\eqref{e:ergod}, ergodicity was expressed by the fact that the
Birkhoff average $\overline{f}^T$ of a test function $f$ on $S^*M$ converges $\mu_L$-almost everywhere towards to
phase space average $\mu_L(\chi)$ when $T\to\infty$. By dominated
convergence, this implies the convergence in the $L^2$ sense, which
amounts to the Von Neumann version of ergodicity, valid for any test
function $f\in L^2(S^*M,\mu_L)$:
\begin{equation}\label{e:vonN}
\lim_{T\to\infty}\int_{S^*M}\big|\overline{f}^T - \mu_L(f)\big|^2\,d\mu_L =0\,.
\end{equation}
We may apply this convergence to the quantum variance
\eqref{e:qu-variance}, taking into account that the LHS is independent
of the time $T$. First fixing $T$ large enough for the classical
variance to be small, then choosing $\hbar$ small for the remainder
$\cO_T(\hbar)$ to be small as well, we find that the quantum variance for $\Oph(\chi)$
vanishes in the semiclassical limit:
\begin{equation}\label{e:QV1}
\lim_{\hbar\to 0}\Var_\hbar([\alpha,\beta]) = 0\,.
\end{equation}
By the Bienaym\'e-Chebychev inequality, this decay of the quantum
variance implies that, when $\hbar\to 0$, the huge majority of the
matrix elements $(\la u_j,\Oph(\chi)u_j\ra)_{j\in J_\hbar}$ will be asymptotically close to $\mu_L(\chi)$.
More precisely, considering a sequence $\hbar^k\to 0$, we can therefore extract
subsets $\tJ_{\hbar^k}\subset J_{\hbar^k}$ with
\begin{equation}
\frac{\# \tJ_{\hbar^k}}{\# J_{\hbar^k}} \xrightarrow{k\to\infty} 1\qquad
\text{(subsets of asymptotic density $1$)},
\end{equation}
such that the matrix
elements with $j\in \tJ_{\hbar^k}$ are close to the classical average:
$$
\lim_{k\to\infty} \max \big\{ |\la u_j,\Op_{\hbar^k}(\chi)u_j\ra -
\mu_L(\chi)|,\ j\in \tJ_{\hbar^k}\big\} = 0\,.
$$
And since we have chosen the interval $[\alpha,\beta]$ so that the
equality \eqref{e:h-hn2} holds throughout $j\in J_\hbar$, we have actually
\begin{equation}\label{e:QEchi}
\lim_{k\to\infty} \max \big\{ |\la u_j,\Op_{\hbar_j}(\chi)u_j\ra -
\mu_L(\chi)|,\ j\in \tJ_{\hbar^k}\big\} = 0\,.
\end{equation}

\subsubsection{Extracting a global subsequence}

So far we have extracted subsets $(\tJ_{\hbar^k})_{k\geq 1}$ of the index sets
$(J_{\hbar^k})_{k\geq 1}$ such as to realize the convergence
\eqref{e:QEchi} for the specific test function $\chi$.

We now want to show that one can extract density one subsequences,
such that the convergence \eqref{e:QEchi} holds for {\it any} function
$f\in C^\infty_c(T^*M)$.
To do so, we may first restrict ourselves to functions constructed
as locally radial extension of functions $\tchi\in C^\infty(S^*M)$, as is
explained in \eqref{e:radial-ext}. 

We start with a remark on the pseudodifferential calculus on $M$. The
Calder\'on-Vaillancourt theorem tells us that the operator norm
$\|\Oph(\chi)\|_{L^2\to L^2}$ is
controlled by finitely many derivatives of $\chi\in C^\infty_c(T^*M)$: 
$$
\exists C_M>0,\qquad \forall\chi\in C^\infty_c(T^*M),\quad \forall \hbar\in ]0,1],\qquad \|\Oph(\chi)\|\leq C_M \|\chi\|_{C^L(T^*M)}\,.
$$
We will then use the fact that, on the compact manifold $S^*M$, the
space $C^\infty(S^*M)$ is {\it separable}: there exists a sequence
$(\tchi_n)_{n\geq 1}$ which is dense in
$C^\infty(S^*M)$, meaning that for any $\tchi\in C^\infty(S^*M)$,
there exists a subsequence $(\tchi_{n_k})$ such that for any $L\in
\IN$, the norms
$\|\tchi_{n_k}-\tchi\|_{C^L}\xrightarrow{k\to\infty} 0$.

Using the procedure of Eq.~\eqref{e:radial-ext}, the
functions $\tchi_n$ can be lifted to locally radial functions $chi_n$, using the
same radial function $\psi$. As a result, the $(\chi_n)_{n\geq 1}$
form a sequence in $C_c^\infty(T^*M)$.

Let us now apply the extraction procedure for the test
function $\chi_1$. This leads to a sequence
$(\tJ_{\hbar^k}^{(1)})_{k\geq 1}$ of subsets of $J_{\hbar^k}$ and
asymptotic density $1$. We then consider the second observable
$\chi_2$. The vanishing of the variance for $\Oph(\chi_2)$ allows to extract 
subsets $\tJ_{\hbar^k}^{(2)}\subset \tJ_{\hbar^k}^{(1)}$ of asymptotic
density one when $k\to\infty$, satisfying \eqref{e:QEchi}. We may proceed iteratively,
constructing subsets $\tJ_{\hbar^k}^{(n)}\subset
\tJ_{\hbar^k}^{(n-1)}$, such that $(\tJ_{\hbar^k}^{(n)})_k$ have of
asymptotic density one inside $J_{\hbar^k}$ when $k\to\infty$:
$$
\lim_{k\to\infty} \frac{\# \tJ_{\hbar^k}^{(n)}}{\# J_{\hbar^k}} = 1.
$$
For each $n\geq 1$ and $p\geq 1$, whe deduce the existence of indices
$k(n,p)\in\IN^*$ such that
$$
\forall k\geq k(n,p),\qquad  \frac{\#
  \tJ_{\hbar^k}^{(n)}}{\# J_{\hbar^k}} \geq 1-\frac{1}{p}.
$$
The double sequence $k(n,p)$ can naturally be chosen to be increasing w.r.t. $p$; we can easily
modify it such as to be strictly increasing w.r.t $n$, e.g. by taking
$$
\tk(n,p) = \max \{k(1,p),\ldots,k(n,p)\} + n\,.
$$
We then proceed by a diagonal extraction from the sequences $(\tJ_{\hbar^k}^{(n)})_{n,k}$, taking the sets
$$
\tJ_k \defeq \tJ_{\hbar_k}^{(n)}\quad \text{in the interval} \quad \tk(n,n) \leq k < \tk(n+1,n+1)\,.
$$
The subsets $(\tJ_k\subset J_{\hbar_k})_{k\geq 1}$ enjoy the following
properties:
\begin{enumerate}
\item The fact that $k\geq \tk(n,n)$ in the above interval shows that $\frac{\#\tJ_k}{\#J_{\hbar^k}} \geq
1-\frac{1}{n}$. As a result, the sequence $(\tJ_k)_{k\geq 1}$ has
asymptotic density $1$ inside the $J_{\hbar_k}$. 
\item For a given
$n\geq 1$, we have $\tJ_k\subset \tJ^{(n)}_{\hbar_k}$ for all $k\geq
k(n,n)$; as a result, the matrix elements corresponding to elements
in $\tJ_k$ satisfy
$$
\forall n\in\IN^*,\qquad\lim_{k\to\infty} \max \{ |\la u_j,\Op_{\hbar_j}(\chi_n)u_j\ra -
\mu_L(\chi_n)|,\ j\in \tJ_{k}\} = 0\,.
$$
\end{enumerate}
Our final task to prove Thm~\ref{thm:QE2} is to show that the above
convergence holds for any compactly supported test function on $T^*M$.

Let us first start with arbitrary locally radial functions of the type
\eqref{e:radial-ext}. Namely,
let us pick an arbitrary $\tchi\in C^\infty_c(S^*M)$, and extend it
into $\check\chi\in C^\infty_c(T^*M)$ as in \eqref{e:radial-ext}. We can
find a subsequence $(\tchi_{n_l})_l$ from our dense sequence,
converging to $\tchi$ in the $C^\infty(S^*M)$ sense. From there, the radial
lifts $(\check\chi_{n_l})_l$ converge (in the $C^\infty(T^*M)$ topology) to the radial lift
$\check\chi$ of $\tchi$. As a consequence, we have 
$$
\mu_L(\check\chi_{n_l})=\mu_L(\chi_{n_l})\xrightarrow{l\to\infty} \mu_L(\check\chi)=\mu_L(\chi),\qquad
\text{and}\quad
\|\Oph(\check\chi_{n_l})-\Oph(\check\chi)\|\xrightarrow{l\to\infty} 0,\quad
\text{uniformly for $h\in (0, 1]$}
$$
It is then obvious that 
$$
\lim_{k\to\infty} \max \{ |\la u_j,\Op_{\hbar_j}(\check\chi)u_j\ra -
\mu_L(\check\chi)|,\ j\in \tJ_{k}\} = 0\,.
$$
Finally, let us consider an arbitrary test function $\chi\in
C^\infty_c(T*M)$. We define its restriction $\tchi=\chi_{\restriction
  S^*M}$, and the radial lift of this restriction, $\check\chi \in
C^\infty_c(T*M)$. Since $\mu_L$ is supported on $S^*M$, the above
shows that
$$
\lim_{k\to\infty} \max \{ |\la u_j,\Op_{\hbar_j}(\check\chi)u_j\ra -
\mu_L(\chi)|,\ j\in \tJ_{k}\} = 0\,.
$$
Our last task will consist in comparing the matrix elements $\la
u_j,\Op_{\hbar_j}(\check\chi)u_j\ra$ and $\la
u_j,\Op_{\hbar_j}(\chi)u_j\ra$. This is done by noticing that the
smooth function $\check\chi - \chi$ vanishes on $S^*M$, and that this
manifold is the noncritical vanishing locus of the function
$H(x,\xi)-1/2$. Standard differential calculus then shows that one can
factorize $\check\chi - \chi$ into
$$
\check\chi - \chi = \theta (H-1/2),\quad \text{for a function}\ \theta\in C^\infty_c(T^*M)\,.
$$
The pseudodifferential calculus allows to factorize as well the
corresponding operators at the principal level:
$$
\Oph(\check\chi)-\Oph(\chi) = \Oph(\theta)
(H_\hbar-1/2)+R_\hbar\,,\quad \text{with}\quad \|R_\hbar\|=\cO(\hbar)
$$
Since $(H_{\hbar_j}-1/2)u_j=0$, the corresponding matrix elements thus satisfy
$$
\la u_j,\Op_{\hbar_j}(\check\chi)u_j\ra = \la
u_j,\Op_{\hbar_j}(\chi)u_j\ra + \cO(\hbar_j)\,,
$$
so that our arbitrary test function $\chi$ satisfies the required limit:
$$
\lim_{k\to\infty} \max \{ |\la u_j,\Op_{\hbar_j}(\chi)u_j\ra -
\mu_L(\chi)|,\ j\in \tJ_{k}\} = 0\,.
$$

\subsubsection{From energy intervals to the full spectrum}

This proves a form of quantum ergodicity when focussing on
the sequence of index sets $J_{\hbar_k}([\alpha,\beta])$, with some
arbitrary sequence
$\hbar_k\searrow 0$.
From the definition \eqref{e:Jh} of these index
sets, we see that the above results allows to scan the full spectrum
of the laplacian. We had chosen our interval
$[\alpha,\beta]\subset [1/8,2]$, so if $\hbar\ll 1$ the interval
$J_\hbar([\alpha,\beta])$ does not contain the bottom of the
spectrum.
Yet,  by choosing the sequence of Planck parameters $\hbar_k$
appropriately, we can make sure that the sets $(J_{\hbar_k})_{k\geq 1}$ form a
partition of the full index set $\IN^*$ (omitting only the zero eigenvalue):
$\IN^*=\bigsqcup_k J_{\hbar_k}.$ By taking then
$$\cS\defeq \bigsqcup_k J_{k},$$
we obtain a subsequence
$\cS\subset\IN^*$ which is itself of asymptotic density 1, and for which
$$
\forall \chi\in C^\infty_C(T^*M),\qquad
\lim_{\cS\ni j\to\infty} \la u_j,\Op_{\hbar_j}(\chi)u_j\ra =
\mu_L(\chi)\,.
$$
This is exactly the statement made in Theorem~\ref{thm:QE2}.

\subsection{Adapting the proof to the cavity setting}

The quantum ergodicity theorem was extended to manifolds with
piecewise smooth boundaries in \cite{ZZ96}. Which ingredients of the
proof need to be modified?

For a Euclidean domain, the counting of eigenvalues of $\Delta_\Omega$
started with the work of 
H.Weyl in 1911, hence long before the invention of pseudodifferential
operators. Weyl used variational methods (min-max theorem and
Dirichlet-Neumann bracketing) to prove the main asymptotics. Later
proofs made use of heat kernel asymptotics on the
diagonal, $e^{t\Delta_\Omega}(x,x)$, to show a {\it local Weyl's law}, that is, for a fixed test function $f\in C(M)$,  the
asymptotics for the sum
$$
\sum_{\lambda_j\leq \Lambda} \int_M f(x) |u_j(x)|^2\,dx = \tr\big(f \bbbone_{[0,\Lambda^2]}(-\Delta)\big) = 
\frac{\Omega_d}{(2\pi)^d}\Lambda^d \int_M f(x)\,dx + \cO(\Lambda^{d-1}),
$$
when $\Lambda\to\infty$. Taking $f\equiv 1$
provides the Weyl asymptotics for the counting function. 
In our previous notations, this estimate reads:
\begin{equation}\label{e:localWL}
\sum_{j\in J_\hbar} \int_M f(x) |u_j(x)|^2\,dx = \frac{\Omega_d}{(2\pi\hbar)^d}
((2\beta)^{d/2}-(2\alpha)^{d/2}) \int_M f(x)\,dx + \cO(\hbar^{1-d}).
\end{equation}

In order to test the eigenmodes $u_j$ also in Fourier space, one needs
to employ pseudodifferential operators, but, as explained in
section~\ref{s:setting-cavity}, one needs to avoid the operator to
touch the boundary. Although \cite{ZZ96} were using polyhomogeoneous
pseudodifferential operators, one can also adapt their proof to
operators of the type $\Oph(\chi)$, with $\chi\in
C^\infty_c(T^*\oOmega)$. Instead of using the Schr\"odinger
propagator $e^{-itH_\hbar/\hbar}$, the authors rather use the
half-wave propagator $e^{-it\sqrt{-\Delta}}$, which quantizes the
broken geodesic flow generated by $p(x,\xi)=|\xi|$ (the difference
with the flow $\Phi^t$ generated by $H=|\xi|^2/2$ is that the speed
along the geodesic does not depend on $|\xi|$).

A trace formula for
$\tr \big(f(H_\hbar)\Oph(\chi)\big)$ similar with the one in \eqref{e:GWL1} was
obtained from estimates on the {\it generalized wave trace}
$\tr\big(\Oph(\chi)\,e^{-it\sqrt{-\Delta}}\big)$ by H\"ormander.
It leads to a generalized Weyl's law identical with \eqref{e:generalizedWL},
but restricted to test functions
$\chi$ supported away from $\partial\Omega$.
In particular, this type of
estimate will not detect a potential positive fraction of the $L^2$
mass asymptotically concentrated near $\partial\Omega$. 

The step involving time evolution up to arbitrary large times $T>0$ is
more delicate. One has to remove from the phase space the ``bad points''
$(x,\xi)\in S^*\oOmega$, whose
future ray $\Phi^t(x,\xi)$ hits a singular point of $\partial\Omega$ before time
$T$ (and cannot be extended beyond that time), or hits
$\partial\Omega$ tangentially at some point $t\in [0,T]$.
For a given time $T>0$, all those bad points make up a ``bad set''
$B_T\subset S^*\Omega$ of zero Liouville measure. One removes these bad points by setting up a
cutoff $\psi_{T,\vareps}\in C^\infty(S^*\oOmega)$ equal to unity away from a small
neighbourhood of $B_T\cup S^*_{\partial\Omega}\Omega$. Since the
latter set has zero measure
ask that $\int (1-\psi_{T,\vareps})\,d\mu_L$ be arbitrary small.

For a given test function $\chi\in C^\infty_c(S^*{\oOmega})$,
the product $\chi_{T,\vareps}\defeq \psi_{T,\vareps}\, \chi$ will be smooth, its
transport through the geodesic flow is well-defined up to time $T$,
including nontangential bounces on the regular points of
$\partial\Omega$. The support of the transported function
$\chi_{T,\vareps}\circ\Phi^t$ will in general intersect
$\partial\Omega$, so a direct quantization of
$\chi_{T,\vareps}\circ\Phi^t$ is problematic. To avoid this
problem, one uses a spatial cutoff function $\psi_\vareps\in C^\infty_c({\oOmega})$,
equal to unity except in an $\vareps$-neighbourhood of
$\partial\Omega$. The Egorov theorem adapted to the cavity then takes the following form:
$$
\forall t\in [0,T],\qquad
\psi_\vareps U^t\Oph(\chi_{T,\vareps}) U_\hbar^t\psi_{\vareps} =
\Oph((\psi_\vareps)^2 (\chi_{T,\vareps}\circ\Phi^t))+\cO_t(\hbar)\,,
$$
where symbols on both sides vanish near $\partial\Omega$.
Performing the average over time, this leads to the approximate
equality between
quantum and classical time averages:
\begin{equation}\label{e:Egorov2}
\psi_\vareps \overline{\Oph(\chi_{T,\vareps})}^T \psi_{\vareps} = \Oph(\psi_\vareps^2\,  \overline{\chi_{T,\vareps}})+\cO_T(\hbar)\,.
\end{equation}
We now perform the same manipulations as in Section~\ref{s:variance}, but must
remain careful of removing the neighbourhood of the boundaries, that is, applying the above operators to the truncated eigenmodes $\psi_\vareps u_j$.
Using the Egorov property \ref{e:Egorov2}, we obtain:
$$
\la \psi_\vareps u_j, \overline{\Oph(\chi_{T,\vareps})}^T
\psi_{\vareps}u_j\ra = \la u_j,\Oph(\psi_\vareps^2\, 
\overline{\chi_{T,\vareps}})u_j\ra +\cO_T(\hbar)
$$
We can use this equality in order to estimate the
quantum variance:
\begin{align}\label{e:Qvar2}
\sum_{j\in J_\hbar}\Big| \la \psi_\vareps u_j, \big(\Oph(\chi_{T,\vareps})
-\mu_L(\chi_{T,\vareps})\big) \psi_\vareps u_j\ra \Big|^2 &\leq
\sum_{j\in J_\hbar} \la \psi_\vareps u_j, \Oph(|\overline{\chi_{T,\vareps}}^T -
\mu_L(\chi_{T,\vareps})|^2) \psi_\vareps u_j\ra +\cO_T(\hbar)\\\nonumber
&\leq \sum_{j\in J_\hbar} \la  u_j, \Oph(\psi_\vareps^2\,|\overline{\chi_{T,\vareps}}^T -
\mu_L(\chi_{T,\vareps})|^2)  u_j\ra +\cO_T(\hbar).
\end{align}
The RHS can be estimated through a generalized Weyl's law similar with \eqref{e:GWL-T}:
$$
\sum_{j\in J_\hbar}\Big| \la \psi_\vareps u_j, \big(\Oph(\chi_{T,\vareps})
-\mu_L(\chi_{T,\vareps})\big) \psi_\vareps u_j\ra \Big|^2 \leq
\frac{\Omega_d\vol(M)}{(2\pi\hbar)^d}
((2\beta)^{d/2}-(2\alpha)^{d/2})
\int_{S^*M}\psi_\vareps^2\, |\overline{\chi_{T,\vareps}}^T -
\mu_L(\chi_{T,\vareps})|^2 \,d\mu_L + \cO_T(\hbar^{-d+1})\,.
$$
In the integral on the RHS, we may choose the cutoffs $(1-\psi_{T,\vareps})$ and $(1-\psi_\vareps)$ with sufficiently small supports, such that 
$$
\int_{S^*M}\psi_\vareps^2\, |\overline{\chi_{T,\vareps}}^T - \mu_L(\chi_{T,\vareps})|^2 \,d\mu_L  =
\int_{S^*M} |\overline{\chi}^T - \mu_L(\chi)|^2 \,d\mu_L +\cO(\vareps).
$$
The ergodicity assumption ensures that, for $T$ large enough, the second integral is smaller than $\vareps$.
For this choice of $T$, and the choice of the cutoffs $\psi_{T,\vareps}$, $\psi_\vareps$, one thus obtains the estimate
\begin{equation}\label{e:est1}
\sum_{j\in J_\hbar}\Big| \la \psi_\vareps u_j, \big(\Oph(\chi_{T,\vareps})
-\mu_L(\chi_{T,\vareps})\big) \psi_\vareps u_j\ra \Big|^2  = \cO(\vareps \hbar^{-d})\,.
\end{equation}
Let us finally take care of the LHS in \eqref{e:Qvar2}, comparing it with the non-cutoff matrix elements
\begin{equation}\label{e:good-sum}
\sum_{j\in J_\hbar}\Big| \la  u_j, \big(\Oph(\chi)
-\mu_L(\chi)\big)  u_j\ra \Big|^2\,.
\end{equation}
Calling $A=\Oph(\chi)
-\mu_L(\chi)$, let us split 
$$
\la u_j,Au_j\ra = \la \psi_\vareps u_j,A\psi_\vareps u_j\ra + \la (1-\psi_\vareps) u_j,A\psi_\vareps u_j\ra + \la  u_j,A(1-\psi_\vareps) u_j\ra
$$
The square of this sum can be bounded by
$$
3\Big( |\la \psi_\vareps u_j,A\psi_\vareps u_j\ra|^2 + |\la (1-\psi_\vareps) u_j,A\psi_\vareps u_j\ra|^2 + |\la  u_j,A(1-\psi_\vareps) u_j\ra|^2\Big)\,.
$$  
The second and third terms on the RHS are bounded above by $\|A\|^2 \| \|(1-\psi_\vareps) u_j\|^2$, a term which can be
estimated by using the local Weyl's law \eqref{e:localWL} (here it is crucial that this local law holds for the function $(1-\psi_\vareps)^2$,  whose support touches $\partial\Omega$):
$$
 \sum_{j\in J_\hbar} \|(1-\psi_\vareps) u_j\|^2 =  C\hbar^{-d} \int_M (1-\psi_\vareps)^2\,dx + \cO(\hbar^{1-d}). 
$$
We may adjust the cutoff $\psi_\vareps$ such that the integral in the RHS is $\cO(\vareps\hbar^{-d})$.
Altogether, we obtain
$$
\sum_{j\in J_\hbar}\Big| \la  u_j, A  u_j\ra \Big|^2 \leq 3  \sum_{j\in J_\hbar}\Big| \la  \psi_\vareps u_j, A \psi_\vareps u_j\ra \Big|^2 + \cO(\vareps\hbar^{-d})\,.
$$
There now remains to replace the operator $A$ associated with the function $\chi$, by the operator $A_{T,\vareps}$ associated with the cutoff function $\chi_{T,\vareps}$.  The same type of simple manipulations yields
\begin{align*}
3\sum_{j\in J_\hbar}\Big| \la  \psi_\vareps u_j, A \psi_\vareps u_j\ra \Big|^2 &= 3\sum_{j\in J_\hbar}\Big| 
\la  \psi_\vareps u_j, A_{T,\vareps} \psi_\vareps u_j\ra + 
\la  \psi_\vareps u_j, (A_{T,\vareps}-A) \psi_\vareps u_j\ra  \Big|^2\\
&\leq 6 \sum_{j\in J_\hbar} \Big| \la  \psi_\vareps u_j, A_{T,\vareps} \psi_\vareps u_j\ra \Big|^2 + 
\Big| \la  \psi_\vareps u_j, (A_{T,\vareps}-A) \psi_\vareps u_j\ra  \Big|^2
\end{align*}
The first sum was estimated in \eqref{e:est1}, while the second one can, again, be estimated by the generalized Weyl's law. The principal symbol of $\psi_\vareps(A_{T,\vareps}-A)\psi_\vareps$ reads 
$\delta \chi = \psi_\vareps^2\Big((1-\psi_{T,\vareps})\chi - \mu_L\big((1-\psi_{T,\vareps})\chi\big) \Big)$, so the Weyl's law gives
$$
\sum_{j\in J_\hbar} \Big| \la   u_j, \psi_\vareps(A_{T,\vareps}-A) \psi_\vareps u_j\ra  \Big|^2 \leq 
C\hbar^{-d} \int  \big| \delta\chi \big|^2\,d\mu_{L}+\cO(\hbar^{-d+1})\,.
$$
Choosing the cutoff $(1-\psi_{T,\vareps})$  with sufficiently small support, we may ensure that the above integral is smaller than $\vareps$. \\
Summing all terms, the original sum \eqref{e:good-sum} is of order $\cO(\vareps\hbar^{-d})+\cO(\hbar^{1-d})$. Dividing by $\#J_\hbar$ and taking $\hbar$ small enough, we get the estimate for the variance
$\Var_\hbar ([\alpha,\beta])= \cO(\vareps)$. Since $\eps>0$ can be chosen arbitrarily, we get the asymptotic vanishing of the variance:
$$
\lim_{\hbar\to 0} \Var_\hbar ([\alpha,\beta]) =0\,.
$$
The rest of the proof after Eq.~\eqref{e:QV1} is identical to the boundaryless case.

\section{The Quantum Unique Ergodicity conjecture}
As already noticed in \cite{CdV85}, the Quantum Ergodicity theorem~\ref{thm:QE2} does not exclude the existence of alternative semiclassical
measures, possibly associated with subsequences $\cS'\subset\IN$ of density 0,
hence corresponding to ``rare'', or ``exceptional'' eigenmodes. This
could be the case, for instance, due to the ``scars'' of periodic
orbits observed in numerical computations: do some sequence of
``scarred eigenmodes'' lead to semiclassical measures $\mu_{sc}$
containing some positive multiple of $\mu_\gamma$, the delta measure
on a closed geodesic $\gamma$? 

In 1994, Rudnick and Sarnak stated a conjecture specific to the manifolds of
negative curvature (on which the geodesic flow is Anosov).
\begin{conj}[Quantum Unique Ergodicity]
Take $(M,g)$ a compact manifold of negative curvature (without
boundary). Then there exist no exceptional eigemodes of the Laplacian:
for any orthogonal eigenbasis $(u_n)_{n\geq 1}$, one can take in Thms~\ref{thm:QE},\ref{thm:QE2}
the full sequence $\cS=\IN$. In other words, for these manifolds there exists a unique
semiclassical measure, which is the Liouville measure: {\it all} high
frequency eigenmodes equidistribute on $S^*M$.

As a consequence, the measures $|u_n(x)|^2\,dg(x)$ weakly converge to
the normalized Lebesgue measure on $M$ when $n\to\infty$.
\end{conj}
The name ``Quantum Unique Ergodicity'' refers to the concept of
``Unique ergodicity'' from classical ergodic theory : a dynamical
system (say, a flow $\Phi^t$ acting on some phase space $\cP$) is said to be uniquely ergodic if it admits a
unique invariant probability measure $\mu$. This property also says that for any initial point
$x\in\cP$, the Birkhoff averages $\frac{1}{T}\int_0^T
f(\Phi^t(x))\,dt$ converges to the ergodic average $\int_{\cP}
f\,d\mu$ when $t\to\infty$. Hence, all orbits of the flow have the
same behaviour for long times.

An example of a uniquely ergodic system is given by
the translation flow on a torus $\IT^2$, $\Phi^t(x)=x+t\omega\bmod
\IZ^2$, provided the two components of the vector $\omega$ are not
commensurate. Indeed, the long segments $\{x+t\omega\bmod\IZ^2,t\in [0,T]\}$
become denser and denser on $\IT^2$ when $T\to\infty$, and they do so
quantitatively, such as to cover the torus evenly. An easy way to
prove this convergence is to decompose the test function $f$ into Fourier modes $e_k(x)=\exp(i
k\cdot x)$, $k\in (2\pi \IZ)^2$.
Such a translation flow is ergodic, but it is not mixing (a localized
cloud of particles will remain a localized cloud under the evolution),
and certainly not hyperbolic.

Certain uniquely ergodic systems (namely {\it discrete time} translations on
the 2-dimensional torus) have been quantized into {\it quantum maps}, that is
$\hbar$-dependent propagators $U_\hbar$ acting on some ad hoc
quantum Hilbert space \cite{BDB96}. The same questions of phase space localization
vs. localization of the eigenmodes of $U_\hbar$ can be
addressed, as well as the definition of semiclassical measures. The
unique ergodicity of the classical system then naturally leads to a unique
semiclassical measure, namely the Lebesgue measure on $\IT^2$.

On the opposite, the above Quantum Unique Ergodicity conjecture addresses the more chaotic flows given by geodesic
flows on negative curvature manifolds. Such flows possess infinitely
many invariant measures (e.g. delta measures on each periodic
geodesic, or measures construct by adding such delta measures), hence they are not uniquely ergodic. Yet, the
conjecture states that the high frequency waves ``select'' one
particular invariant measure, namely the Liouville measure, and
discard all other invariant measures.

\subsection{Arithmetic QUE}
The conjecture remains open, even in the case of manifolds of
constant negative curvature. Certain compact surfaces of
constant negative curvature $M$ (a.k.a. congruence hyperbolic surfaces) feature
arithmetic symmetries, embodied in the existence of an infinite family
of selfajdoint {\it Hecke
operators} $(T_n)_{n\geq 1}$ acting on $L^2(X)$, which commute with
each other and with the Laplacian on $M$. These Hecke operators can be
viewed as generating arithmetic symmetries of the manifold. It then
makes sense to diagonalize them simultaneously with the Laplacian,
that is consider specifically our eigenmodes $(u_n)_{n\in\IN}$ to be
joint eigenmodes of the Laplacian and all the Hecke operators.
In those conditions, Lindenstrauss \cite{Lin06} proved that the full
sequence $(u_n)_{n\in\IN}$ satisfies QUE (due to this arithmetic
structure, one speaks of {\it arithmetic QUE}). In a later work,
Brooks and Lindenstrauss \cite{BrL14} showed that one only needs ot
assume that $(u_n)_n$ are eigenmodes of the Laplacian and one Hecke
operator. The proofs of arithmetic QUE is very different from that of
Quantum Ergodicity. It relies on positive entropy estimates for the
semiclassical measures (using the Hecke symmetries), as well as rigidity theorems on the 
measures which are both invariant w.r.t. the geodesic flow, and
{\it recurrent} w.r.t. the Hecke correspondences. 

Is this requirement to be joint eigenmodes necessary? There is a
conjecture about the fact that the eigenvalues of the Laplacian
on such congruence surfaces should be simple \cite{Sar11}. If this is
the case, all eigenmodes of the surface would then automatically be joint eigenmodes
of all the Hecke operators, and hence be asymptotically
equidistributed.
This spectral simplicity conjecture is
confirmed by numerical computations of the spectrum of some congruence
surfaces, but proving it mathematically seems at
present even more distant than the QUE conjecture.

\subsection{Counterexample to QUE on a chaotic Euclidean billiard}\label{s:Hassell}

The QUE conjecture is still open on manifolds of negative
curvature, but it does not generalize to any ergodic or mixing
flow. Indeed, as mentioned in the introduction, there are Euclidean
billiards proven to be ergodic and mixing w.r.t. the Lebesgue measure,
but in which some orbits are marginally stable (that is, have zero
Lyapunov exponent). This is the case of the stadium billiard \cite{Bun79}, featured
in Fig.~\ref{f:3modes}. The central rectangular part in this billiard
allows for a 1-parameter family of vertical {\it bouncing ball} orbits hitting
the horizontal boundaries, hence which do never enter the ``circular
wings'' of the billiard. Numerical computations exhibit sequences of
{\it bouncing ball eigenmodes} with very pronounced concentration
inside the rectangle, and in phase space along those bouncing ball
orbits (see the right plot in Fig.~\ref{f:3modes}). 

It was known that, using this central rectangle (say, $[0,L_1]\times[0,L_2]$), one can cook up
quasimodes as follows:
\begin{equation}\label{e:v_k}
v_k(x) = f(x_1)\,\sin(\pi k
x_2/L_2)=\frac{f(x_1)}{2i}\big(e^{i\tlambda_k x_2} - e^{-i\tlambda_k
  x_2}\big),\quad
k\in\IN^*,\quad f\in C^\infty_c(]0,L_1[),
\end{equation}
where we defined $\tilde\lambda_k=\pi k/L_2$. We also choose to
normalize $f$ s.t. $\|v_k\|=1$.
In the large-$k$ regime, this quasimode satisfies
\begin{equation}\label{e:quasimode}
(\Delta +\tilde\lambda_k^2)v_k = f''(x_1)\sin(\tlambda_k x_2) = \cO(1)_{L^2},,\quad k\to\infty.
\end{equation}
We say that $v_k$ is a
quasimode of central frequency $\tlambda_k$ and
error $\cO(1)$. The spectral theory of selfadjoint operators tells us
that there is a constant $C>0$ (independent of $k$) such that the
spectral decomposition of $v_k$ into true eigenmodes $u_n$ has a
positive weight $W>0$ in the spectral interval
$[\tlambda_k^2-C,\tlambda_k^2+C]$.
In particular, this interval must contain eigenvalues of
$-\Delta$. But how many of them?

Weyl's counting law for the Laplacian implies that the number $N(\Lambda)$ of eigenvalues of $-\Delta$ in an
interval $[\lambda^2-C,\lambda+C]$ should, {\it on average}, satisfy
$N(\lambda)=\cO(1)$ when $\lambda\to\infty$. But this average phenomenon
is not guaranteed to hold uniformly for all $\lambda$.

If by chance our specific intervals $[\tlambda_k^2-C,\tlambda_k^2+C]$
satisfy $N(\tlambda_k)\leq N$ when $k\to\infty$, then the
decomposition of $v_k$ into the $N(\tlambda_k)$ eigenmodes $u_n$ has a
weight $\geq W$; as a result, there is at least one of such
eigenstates $u_n$ for which the weight
\begin{equation}\label{e:overlap}
|\la v_k,u_n\ra|^2\geq \frac{W}{N(\tlambda_k)}\geq
\frac{W}{N},\qquad \text{uniformly w.r.t. $k$.}
\end{equation}
Now, adjusting $\hbar_k=\tlambda_k^{-1}$ as for eigenmodes, the quasimode $v_k$ of \eqref{e:v_k} is the sum of two truncated
plane waves with momenta $\xi_\pm=(0,\pm 1)$. One can easily check
that the sequence of
quasimodes $(v_k)_{k\geq 1}$ is associated with the single measure
$$
\mu_{qm}=\frac{1}{4}|f(x_1)|^2\bbbone_{[0,L_2]}(x_2)\big(\delta(\xi-\xi_+)+\delta(\xi-\xi_-)),
$$
The fact that the eigenstate $u_n=u_{n(k)}$ has a positive overlap
with $v_k$
implies that any semiclassical measure $\mu_{sc}$ extracted from the sequence
$(u_{n(k)})_k$ must contain some positive multiple of $\mu_{qm}$:
$\mu_{sc}\geq W/N \mu_{qm}$. Since $\mu_{qm}$ is singular w.r.t. the
Liouville measure, $\mu_{sc}$ must be different from $\mu_{L}$, so the
subsequence $(u_{n(k)})_k$ is indeed exceptional.

This quasimode construction had been noticed as early as 1988 by
O'Connor and Heller \cite{OCH88}, then by Zelditch \cite{Zel04}. What
remained to show is the
existence of an infinite sequence of intervals
$([\tlambda_k^2-C,\tlambda_k+C])_{k\in\cS}$, each one containing less
than $N$ eigenvalues. This last delicate step was achieved by A.Hassell, at the
cost of considering a 1-parameter family $(S_L)_{L\in [1,2]}$ of stadium billiards,
indexed by the length $L=L_1$ of the central rectangle (keeping its height
unchanged).
\begin{theorem}\cite{Hass10}
For every $\vareps> 0$, there exists a subset $B_\vareps\subset[1,2]$ of measure at
least $1-4\vareps$, and a strictly positive constant $w(\vareps)>0$ with the following property. For
every $L\in B_\vareps$, there exists a semiclassical measure
$\mu^{(L)}_{sc}$ of the billiard $S_L$, which puts a mass at least
$m(\vareps)$ on the family of bouncing ball trajectories of $S_L$.
\end{theorem}
The proof proceeds by analyzing the variation of the eigenfrequencies
$\lambda_n^{(L)}$ when varying the length $L$ of the
central rectangle. In the 2-dimensional cavity $S_L$, Weyl's law
implies that $\lambda_n^2\sim \frac{4\pi n}{
\vol(S_L)}$ when $n\to\infty$, showing that, {\it on
average}, the eigenfrequencies decrease linearly in $L$ when one
increases $L$ (hence the area of $S_L$). On the other hand, the
variation of an individual eigenfrequency $\lambda_n$ depends on the distribution of the
eigenmode $u_n$ along the circular parts of the boundary, which is also
related with the phase space measure $\mu_n$.
Cleverly analyzing these various constraints allows to control, for most
billiard shapes $S_L$, the number of
eigenvalues in an infinite set of intervals
$[\tlambda_k^2-C,\tlambda_k+C]$.\qed

\section{Towards QUE: constraints on localization for all eigenstates}

Attempts to prove Quantum Unique Ergodicity should proceed by investigating individual eigenstates, instead of summing over a large bunch of them, as was done in the proof of Quantum Ergodicity. Some of these attempts, performed in the setting of compact manifolds of negative curvature without boundaries, lead to nontrivial constraints on the possible semiclassical measures. Grossly speaking, they show that, even if some semiclassical measures are different from the Liouville measure (which would breach QUE), these semiclassical measures cannot be too localized, but should satisfy some minimal amount of delocalization.

We will list two of these results, to which the author has partially contributed. Both results proceed by quantizing a finite partition of unity of the energy shell $S^*M=H^{-1}(1/2)$, and refining this partition through the quantum evolution.

\subsection{Refined classical and quantum partitions}
In classical mechanics, one can use sharp partitions of $S^*M$. However, at the quantum level we prefer to use smoothened versions of such partitions. 
\begin{definition}\label{def:partition}
A classical partition of $S^*M$ is a partition $S^*M=\bigsqcup_{k=1}^K\Omega_k$, where each $\Omega_k$ has nonempty interior, and has piecewise smooth boundaries. For $\vareps>0$ small, we may consider the $\eps$-retractions $\Omega_{k-}$ and $\vareps$-neighbourhoods $\Omega_{k+}$ of $\Omega_k$ inside $S^*M$. We thicken these into subsets of $T^*M$,
$\Omega_{k-\vareps}=\bigcup_{|r-1|\leq \vareps/2} r\Omega_{k-}$, $\Omega_{k+\vareps}=\bigcup_{|r-1|\leq \vareps} r\Omega_{k+}$.

A smoothed partition of $S^*M$ is a family of functions $a_k\in C^\infty_c(T^*M)$, such that $a_k=1$ on $\Omega_{k-\vareps}$, $\supp a_k\subset \Omega_{k+\vareps}$, and 
\begin{equation}\label{e:sm-cl}
\sum_{k=1}^K a_k(x,\xi)=1\quad \text{on}\quad \bigcup_{|r-1|\leq \vareps/2} S_r^*M\,.
\end{equation}
A quantum smoothed partition of unity is the family of operators $(A_{k}=\Oph(a_k))_{k=1,\ldots,K}$, such that 
\begin{equation}\label{e:sm-qu}
\sum_{k=1}^K A_k = Id\quad\text{microlocally near $S^*M$.}
\end{equation}
The latter property means that for any test function $\chi\in C^\infty_c(T^*M)$ supported in the energy layer $\{1-\vareps/2\leq |\xi|\leq 1+\vareps/2\}$, one has 
$$
\big(\sum_{k=1}^K A_k\big)\Oph(\chi) =\Oph(\chi) + \cO(\hbar^\infty)\,.
$$
\end{definition}
As a consequence, for an eigenstate $u_j$ and taking $\hbar=\hbar_j$, the partition of unity applies to the eigenstate:
\begin{equation}\label{e:qu-partition}
\big(\sum_{k=1}^K A_k\big)\,u_j = u_j +\cO(\hbar_j^\infty)\,.
\end{equation}
The role of each $A_j$ is to capture the part of $u_j$ microlocalized in $\Omega_{k+\vareps}$. We will not pay attention to the slight difference between $\Omega_k$ and $\Omega_{k\pm\vareps}$, and will speak of ``the part of $u_j$ microlocalized in $\Omega_k$".

The quantum propagator at integer times will then be used to propagate this partition of unity: we then define the evolved classical and quantum partitions by:
$$
a_k(t)\defeq a_k\circ \Phi^t,\qquad A_k(t)\defeq U_\hbar^{-t}\,A_k\,U_\hbar^t,\quad t\in\IZ\,.
$$
The partitions at different integer times may be superposed, to create {\it refined} partitions. For a time $n\in\IN$, we will use the following notation for symbolic words
$$
\balpha = \alpha_0\cdots\alpha_{n},\qquad \alpha_t\in\{1,\ldots,K\}\,.
$$
The time-$n$ refinement of the partition $S^*M=\bigsqcup\Omega_k$ is the partition into the sets
$$
\Omega_{\balpha}=\Omega_{\alpha_0}\cap \Phi^{-1}(\Omega_{\alpha_1})\cap \Phi^{-2}(\Omega_{\alpha_2})\cdots \Phi^{-n}(\Omega_{\alpha_n})\,,
$$
noting that some of these sets may be empty. Each set $\Omega_{\balpha}$ contains the points sharing the same ``symbolic history" between times $0$ and $n$, with respect to the partition $(\Omega_k)_k$. Namely, the points $(x,\xi)\in \Omega_{\balpha}$ are characterized by sitting in $\Omega_{\alpha_0}$ at the time $t=0$, sitting in $\Omega_{\alpha_1}$ at time $t=1$, etc until the time $t=n$. 
\begin{figure}[ht]
  \begin{center}
  \includegraphics[width=0.9\tw]{
  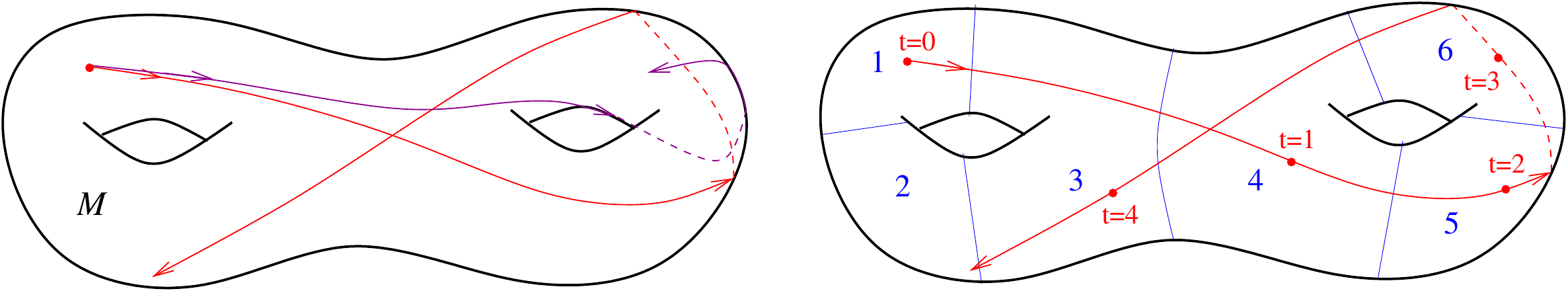}
  \caption{\label{f:hyperb}Left: a pair of orbits issued from the same point of a surface $M$ of negative curvature, departing fast from one another. Right: a partition of $M$ allows to differentiate points according to their ``symbolic history". The indicated point follows the symbolic path $\balpha=14563\ldots$}
  \end{center}
\end{figure}

The smoothened classical and quantum partitions (\ref{e:sm-cl},\ref{e:sm-qu}) are similarly refined into the partitions
\begin{align}
a_{\balpha}&=a_{\alpha_n}(n) \cdots a_{\alpha_1}(1) a_{\alpha_0}\,,\\
A_{\balpha}&=A_{\alpha_n}(n) \cdots A_{\alpha_1}(1) A_{\alpha_0}=U_\hbar^{-n}A_{\alpha_n}U_\hbar^1\cdots U_\hbar^1 A_{\alpha_1}U_\hbar^1A_{\alpha_0}\,.
\label{e:Abalpha}\end{align}
The quantum-classical correspondence and the pseudodifferential calculus show that the two partitions are related to one another:
\begin{equation}\label{e:Q-cl-corresp}
A_k(t)=\Oph(a_k(t))+\cO_t(\hbar)\,,\qquad A_{\balpha}=\Oph(a_{\balpha})+\cO_n(\hbar)\,.
\end{equation}

\subsection{Uniformly hyperbolic dynamics, and the Ehrenfest time}\label{s:Ehren}
The correspondence between classical and quantum refined partitions \eqref{e:Q-cl-corresp} is stated here for a fixed time $n$, sending $\hbar\searrow 0$.  We will however need to consider these quantum refined partitions for times $n$ which explicitly depend on $\hbar$ in a logarithmic way: $n\sim \eta|\log\hbar|$ for some $\eta>0$. It will be crucial that, for some range of $\eta$, the above correspondence will still approximately hold, up to an error $\cO(\hbar^\delta)$ for some $\delta>0$). But beyond a certain threshold $\eta_{th}$, this estimate will break down, essentially because the classical functions $a_k\circ\Phi^t$ will become too singular, in the sense that this function will fluctuate on too short scales. 

Let us come back to the properties of the geodesic flow on manifolds of negative curvature. Due to the negative curvature, all orbits are exponentially unstable. This unstability is formalized by the existence of unstable and stable directions transversely to the flow. More precisely, for each point $\rho=(x,\xi)\in S^*M$,  the linearization of the flow allows to define stable and unstable subspaces $E_s(\rho), E_u(\rho)\subset T_\rho(S^*M)$ with the following properties:
\begin{align*}
\forall v\in E_s(\rho),\ \forall t>0,\qquad \|d\Phi^t v\|&\leq C\,e^{-t\lambda}\|v\|,\\
\forall v\in E_u(\rho),\ \forall t<0,\qquad \|d\Phi^t v\|&\leq C\,e^{-|t|\lambda}\|v\|\,.
\end{align*}
The subspaces $E_{u/s}(\rho)$ are each of dimension $d-1$, they depend continuously on $\rho\in S^*M$ (in general, no better than H\"older continuously), and $d\Phi^t(\rho)$ maps $E_{s/u}(\rho)$ to $E_{s/u}(\Phi^t(\rho))$. 
These subspaces are tangent to the (local) stable and unstable manifolds $W_{u/s}(\rho)\subset S^*M$ which foliate the phase space.

\begin{figure}
\begin{center}
\includegraphics[width=.8\tw]{
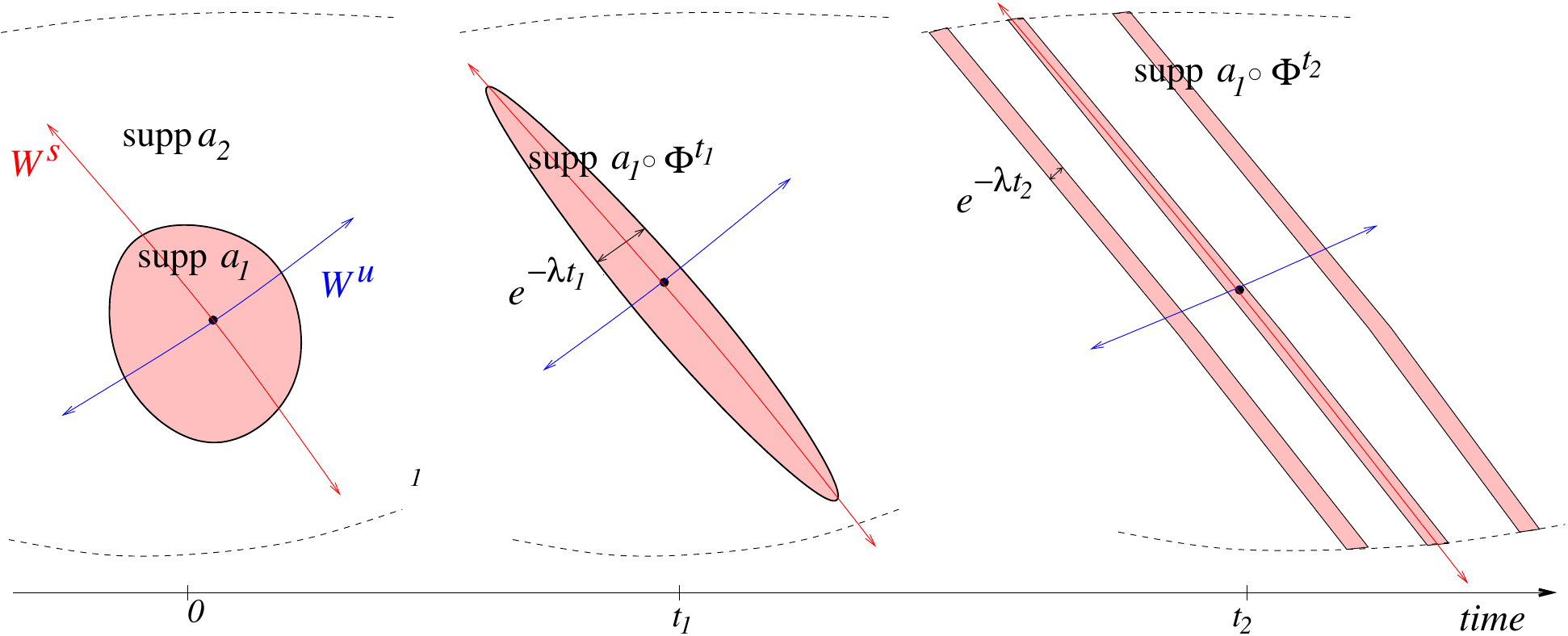}
\caption{\label{f:evol-a}Evolution of a classical cutoff function $a_1$ through the geodesic flow (the pink boxes refer to the supports of the evolved functions). The plots are showing the structures transverse to the flow $\Phi^t$.\label{f:evol-a1}}
\end{center}
\end{figure}

Here the constants $C,\lambda>0$ can be chosen uniformly w.r.t. $\rho$: one speaks of uniform hyperbolicity. The parameter $\lambda$ appearing above can be viewed as the slowest expansion (or contraction) rate. On the opposite, one can define the fastest (or maximal) expansion rate 
$$
\lambda_{\max} = \lim_{t\to\infty}\frac{1}{t}\max_{\rho\in S^*M}\log \|d\Phi^t(\rho)\|\,.
$$
\begin{figure}
  \begin{center}
  \includegraphics[width=0.50\tw]{
  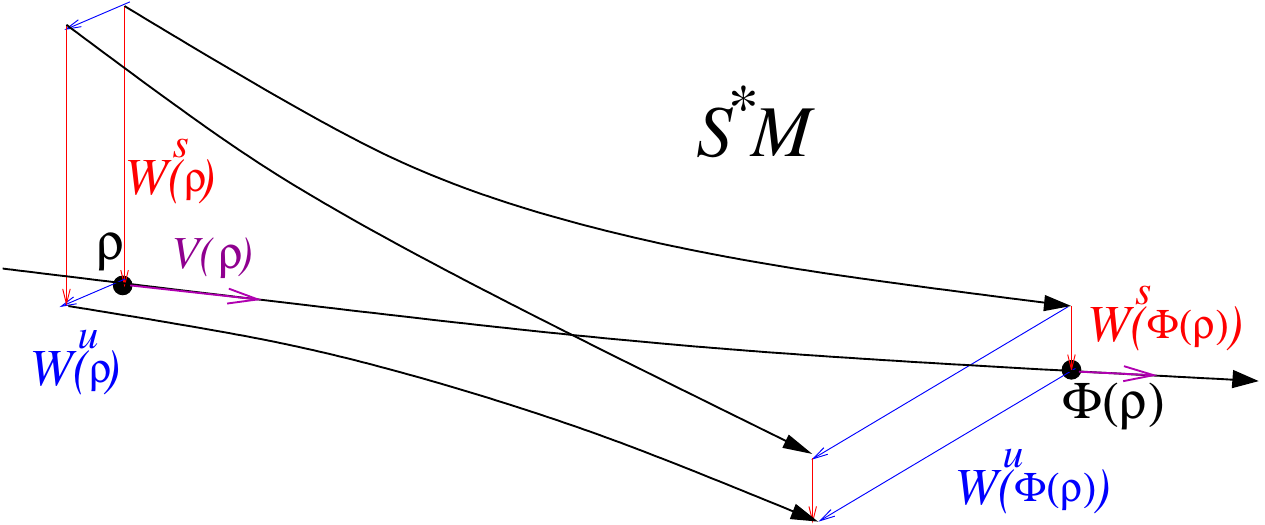}
  \caption{\label{f:Anosov}A phase space picture of a reference orbits $\{\Phi^t(\rho)\}$ and three nearby orbits, two of them along the stable and unstable manifolds of $\rho$.}
  \end{center}
\end{figure}
As a consequence of this exponential unstability, the evolved functions $a_k\circ\Phi^t$ can become very steep when $t$ increases:
$$
\sup_\rho |\partial (a_k\circ\Phi^t)|\leq C\,e^{t\lambda_{\max}},\quad t\to\pm\infty\,.
$$
Equivalently, the function $a_k\circ\Phi^t$ oscillates on scales $\sim e^{-t\lambda_{\max}}$. Figure~\ref{f:evol-a} shows the supports of a evolved functions $a_1\circ\Phi^t$ for positive times: the supports become thin neighbourhoods of some stable manifolds $W^s$, implying that the derivatives of $a\circ\Phi^t$ transversely to those manifolds grow exponentially.

The pseudodifferential calculus and hence the quantum-classical correspondence, will hold as long as the functions do not oscillate on distances smaller than $\hbar^{1/2}$. In view of the above bound, starting from functions $a_k$ with derivatives $\cO(1)$, the shortest time when $a_k\circ\Phi^t$ will oscillate on distances $\sim \hbar^{1/2}$ will be $\frac{1}{2}T_{E}$, where we introduce
\begin{equation}\label{e:Ehren}
\text{the Ehrenfest time}\quad T_E=T_E(\hbar)\defeq\frac{|\log\hbar|}{\lambda_{\max}}\,.
\end{equation}
Hence, as long as $\eta<\frac{1}{2\lambda_{\max}}$, the correspondence principle and pseudodifferential calculus will allow to write
$$
\text{for any word $\balpha$ of length $n\leq \eta|\log\hbar|$},\qquad A_{\balpha} = \Oph(a_{\balpha})+\cO(\hbar^\eps)\,,
$$ 
and the operator $A_{\balpha}$ is a ``good" pseudodifferential operator. In particular, its operator norm 
$\|A_{\balpha}\| =  \sup |a_{\balpha}|+\cO(\hbar^\eps)$.

\subsection{Lower bounds on the Kolmogorov-Sinai entropy}
What are those refined partitions useful for? 
At the classical level, refined partitions $S^*M=\bigsqcup_{|\balpha|=n}\Omega_{\balpha}$ allow to define the {\it Kolmogorov-Sinai entropy} of a flow-invariant probability measure $\mu$. Namely, at each time step $n$, one defines the time-$n$ Shannon entropy of the measure $\mu$ for the refined partition $(\Omega_{\balpha})_{|\balpha|=n}$ by:
$$
H_{n}(\mu) = -\sum_{\balpha} \mu(\Omega_{\balpha})\,\log\big( \mu(\Omega_{\balpha}) \big)\,.
$$
By subadditivity, the limit $\lim_{n\to\infty} \frac{1}{n}H_n(\mu)$
is well-defined. Provided the initial partition $(\Omega_k)_{k}$ has been chosen thin enough, then one can show that the above limit does not depend on this initial choice, it defines the Kolmogorov-Sinal entropy $H_{KS}(\mu)$. The latter has a meaning in information theorey: it characterizes the gain of information acquired by each refinement step, with respect to the measure $\mu$. 

More geometrically, the entropy also provides some information on the concentration of the measure $\mu$. A positive entropy implies some level of delocalization of the measure. For instance, invariant measures $\mu=\sum_i p_i \delta_{\gamma_i}$ supported on countably many closed geodesics $\gamma_i$ all have zero entropy. Similarly, a measure $\mu$ admitting a thin support (in terms, for instance, of Hausdorff dimension) will have a small entropy. 

For any invariant measure $\mu$, the entropy satisfies Ruelle's inequality
$$
H_{KS}(\mu) = \int \log J^u\,d\mu\,,\quad\text{where $J^u(\rho) = |\det d\Phi^1_{\restriction E^u(\rho)}|$}
$$
is the {\it unstable Jacobian}, measuring the exponential growth rate along the unstable subspace. 
This inequality is saturated by the Liouville measure $\mu_L$:
$$
H_{KS}(\mu_L) = \int \log J^u\,d\mu_L\,,\quad\text{in particular, on a manifold of constant curvature $-1$ the value is }H_{KS}(\mu_L) =d-1\,.
$$
On a manifold of variable negative curvature, there can exist measures with even higher entropies than $\mu_L$, like the measure of maximal entropy $\mu_{\max}$. Those measures have full support on $S^*M$. This is why we may view the entropy $H_{KS}(\mu)$ as a sign of the level of delocalization of $\mu$.

With this point of view, the results by Anantharaman \cite{Ana08} and Anantharaman-Nonnenmacher \cite{AN07,AKN09} provide constraints on the possible concentration of semiclassical measures on such manifolds.

\begin{theorem} \cite{Ana08,AN07,AKN09}\label{thm:entropy}
Let $(M,g)$ be a compact manifold of negative sectional curvature. Then there exists $c_M>0$ such that any semiclassical measure $\mu_{sc}$ for the Laplacian on $M$ satifies the lower bounds:
$$
H_{KS}(\mu_{sc}) \geq \max\Big(c_M, \int \log J^u\,d\mu_{sc} - \frac{(d-1)\lambda_{\max}}{2}\Big)\,.
$$
In particular, on a manifold of constant negative curvature $-1$, any semiclassical measure satisfies
$$
H_{KS}(\mu_{sc}) \geq \frac{d-1}{2}\,.
$$
On compact {\it surfaces} of nonpositive curvature, Rivi\`ere showed the cleaner lower bound \cite{Riv10}:
\begin{equation}\label{e:Riv10}
H_{KS}(\mu_{sc}) \geq \frac{1}{2}\int \log J^u\,d\mu_{sc}\,.
\end{equation}
\end{theorem}
The constant $c_M$ is not very explicit, but it is always strictly positive. On the other hand, the second expression on the RHS may become negative if the curvature varies a lot.
This lower bound prevents many invariant measures of low entropy to appear as semiclassical measures. In some sense, stationary wave modes necessarily involve some minimal amount of delocalization of the energy. 
These lower bounds are weaker than the QUE conjecture: the latter would be proved if one could show that $H_{KS}(\mu_{sc}) \geq\int \log J^u\,d\mu_{sc}$, that is removing the factor $1/2$ from \eqref{e:Riv10}.

\subsection{Entropy bounds: sketch of the proof}

We will only give a brief sketch of the proof, more details are given in the original articles and in the Proceedings \cite{Non13}.

Let us assume that a semiclassical measure $\mu_{sc}$ is associated with a certain subsequence of eigenmodes $(u_{j_k})_{k\geq 1}$. To avoid too cumbersome notations, we will call $u_\hbar$ the generic element of this subsequence, where $\hbar=\hbar_{j_k}$.

In order to compute the entropy of $\mu_{sc}$, one needs to estimate the weights $\big(\mu_{sc}(\Omega_{\balpha})\big)_{|\balpha|=n}$, for refinements of large times $n$. Indeed, the Shannon entropy $H_n$ will be large if most of those weights are small, say smaller than $e^{-\beta n}$ for some $\beta>0$. So the goal is to show that most of those weights are small when $n\gg 1$.

Because we prefer to work with smooth functions rather than characteristic functions, we approximate those weights by the integrals
$\int a_{\balpha} d\mu_{sc}$, where $(a_{\balpha})_{|\balpha|=n}$ form the smooth refined partition described above. 
For a fixed word $\balpha$, each such weight will be the limit of 
$$
\mu_{sc}(a_{\balpha}) = \lim_{\hbar\to 0} \la u_\hbar,\Oph(a_{\balpha})u_\hbar\ra = \lim_{\hbar\to 0} \la u_\hbar,A_{\balpha} u_\hbar\ra\,,
$$
where we made use of the quantum-classical correspondence.
However, if we fix $\balpha$ as $\hbar\to 0$, the information we have on the operator $A_{\balpha}$ is that it is a pseudodifferential operator microlocally supported on $\Omega_{\balpha}$, and whose norm can be as large as $\sup|a_{\balpha}|$, namely unity. From this information, we only recover the trivial bound $\mu_{sc}(a_{\balpha}) \leq 1$.

In order to gain some better knowledge, we need to let $n$ increase when $\hbar\to 0$, in a very specific way.  In section~\ref{s:Ehren} we explained that the above quantum-classical correspondence continues to hold if $n$ is as large as $\eta|\log\hbar|$ when $\eta<\frac{1}{2\lambda_{\max}}$. In order to improve our understanding of the entropy, we need to trespass this threshold, that is consider times $n\sim \eta|\log\hbar$ with $\eta>\frac{1}{2\lambda_{\max}}$. For those times, the operator $A_{\balpha}$ is no more a nice operator, because its associated classical function $a_{\balpha}$ is too singular. Yet, a simple time shift conjugates the operator $A_{\balpha}$ to the nicer operator, resp. $a_{\balpha}$ to a nicer function $\tilde a_{\balpha}$:
\begin{align}\label{e:tA}
\tilde A_{\balpha} &= U_\hbar^{-n/2} A_{\balpha} U_\hbar^{-n/2} = A_{\alpha_{n}}(n/2)\cdots A_{\alpha_{n/2+1}}(1) A_{\alpha_{n/2}} A_{\alpha_{n/2-1}}(-1)\cdots A_{\alpha_{0}}(-n/2) \,,\\
\tilde a_{\balpha}&= a_{\balpha}\circ\Phi^{-n/2} = a_{\alpha_{n}}\circ\Phi^{n/2}\cdots a_{\alpha_{n/2}} A_{\alpha_{n/2-1}}\circ\Phi^{-1}\cdots a_{\alpha_{0}}\circ\Phi^{-n/2}\,.\label{e:ta}
\end{align}
$\tilde a_{\balpha}$ is still a ``good function", and $\tilde A_{\balpha}$ a "good pseudodifferential operator", as long as $n/2<\frac{|\log\hbar|}{2\lambda_{\max}}$, hence as long as $n<T_E$.
Since the eigenmode $u_\hbar$ is an eigenmode of the propagator $U_\hbar^t$, this time shift does not modify the matrix element:
$$
\la u_\hbar,A_{\balpha}u_\hbar\ra = \la u_\hbar,\tilde A_{\balpha}u_\hbar\ra \quad\text{as long as }|\balpha| = n<T_E\,.
$$
So we can gain no relevant information on those matrix elements, as long as $n<T_E$. 

\subsubsection*{Trespassing the Ehrenfest time}
However, for times $n>T_E$, even the shifted operator $\tilde A_{\balpha}$ ceases to be a good pseudodifferential operator. On the other hand, for such long times, one is able to prove a {\it dispersion estimate} for the norm of this operator, as long as the elements  $\Omega_k$ of the initial partition have sufficiently small diameters:
\begin{equation}\label{e:disp}
\| A_{\balpha} \| \leq \min \Big(1, C\hbar^{-d/2}\,e^{-\Lambda n/2}\Big)\,,
\end{equation}
where $\Lambda>0$ is the minimal growth rate of the unstable Jacobian when $t\to\infty$. This upper bound is obtained by explicitly computing the evolution of WKB wavepackets of the form $\psi_\hbar(x)=a(x)e^{i\varphi(x)/\hbar}$ through the alternate evolution through $U_\hbar^1$ and the cutoff operators $A_{\alpha_t}$:
$$
\psi_\hbar(n)\defeq A_{\alpha_n}\cdots A_{\alpha_2}U^1_\hbar A_{\alpha_1}U^1_\hbar A_{\alpha_0}\psi_\hbar \,.
$$
The states $\psi_\hbar(n)$ are still of the WKB form, with a precise control on the phase functions and amplitudes. As had been noticed long ago by Tomsovic and Heller \cite{TH91}, the semiclassical description of this evolution does not break down at the Ehrenfest time, but can be followed much longer (at least up to any logarithmic time $C|\log\hbar|$). To compute the norm of the operator, one combines such WKB states $\psi_\hbar$ to construct an arbitrary initial state $v(0)$ microlocalized near the $S^*X$, and adds up those WKB evolutions to deduce a bound on the norm of $v(n)$. 

The small value of the norm $\| A_{\balpha} \| $ will, of course, induce a small value for the matrix element $\la u_\hbar, A_{\balpha} u_\hbar\ra$, which is what we are looking for. Yet, one cannot blindly take such a small upper bound and deduce a lower bound for $H_n(\mu_{sc})$ from it. Indeed, there are several caveats: 

1) the matrix elements $\la u_\hbar, A_{\balpha} u_\hbar\ra$ are not necessarily positive; 

2) even if they were so, the ``quantum" Shannon entropies $H_n$ computed from these matrix elements do not satisfy subadditivity, 

3) these ``quantum entropies" are not directly connected with the ``classical entropies" $H_n(\mu_{sc})$. 

Several tricks are hence necessary to connect the "quantum" Shannon entropies defined from these matrix elements, and the classical entropies $H_n(\mu_{sc})$. One of these tricks is an Entropic Uncertainty Principle, initially discovered in the context of operator algebras, which directly makes use of the dispersion estimate \eqref{e:disp}. This Uncertainty principle allows to get nontrivial lower bound for quantum entropies $H_n$; an approximate subadditivity then has to be proved, which allows to bring lower bounds for these "long times" $H_n$ (on the scale of $n\approx 2 T_E$), to lower bounds for the "short times" entropies ($n\sim \eps T_E$), finally to fixed times $n=\cO(1)$, where one can make contact with the classical entropies $H_n(\mu_{sc})$.

Although this account is very sketchy, the reader should keep in mind that, in order to obtain such nontrivial lower bounds on the delocalization of eigenfunctions, it is necessary to control the evolution beyond the Ehrenfest time, so that the operators $A_{\balpha}$ (and even the symmetrized operators $\tilde A_{\balpha}$) are non longer nice pseudodifferential operators. The same type of ideas will be used to prove the second result on delocalization, which we now present.

\subsection{Full delocalization on Anosov surfaces}

As explained above, the lower bound on the entropy $H_{KS}(\mu_{sc})$ gives some constraint on the possible concentration of the semiclassical measure. Yet, it does not prevent the measure from being supported on a proper invariant subset of $S^*M$. For the chaotic flows of Anosov type, it is indeed relatively easy to construct invariant measures $\mu$ with large enough entropy, but support given by a fractal strict subset of $S^*M$ (necessarily of vanishing Liouville measure). Such measures would then be allowed by the lower bounds on the entropy. 
The result below, which holds only in two dimensions, forbids the existence of such measures among the semiclassical measures, forcing them to be fully supported. 
\begin{theorem}\cite{DJ18,DJN22}
 Let $(M,g)$ be a compact surface of negative curvature. Then any associated semiclassical measure $\mu_{sc}$ has full support $S^*M$. More precisely, for any nonempty open set $\cV\subset S^*M$, there exists a constant $c_{\cV}>0$ such that all semiclassical measures satisfy $\mu_{sc}(\cV)\geq c_{\cV}$.
 
As a consequence, we also have a uniform delocalization of the densities $|u_j(x)|^2$ on $M$: for any nonempty open set $V\subset M$, there exists a constant $c_V>0$ such that all the eigenmodes $u_j$ satisfy 
 $$
 \int_{V}|u_j(x)|^2\,d\vol(x)\geq c_V\,.
 $$
\end{theorem}
We notice that, on any Riemannian manifold (or any domain $\Omega$), the weight $\int_V |u_j(x)|^2\,d\vol(x)$ is never zero, due to the unique continuation property of the eigenfunctions. However, the lower bound generally depends on the eigenfrequency $\lambda_j$, and can be as small as $e^{-c \lambda_j}$. The strength of the lower bound in the above theorem lies in the fact that the constant $c_V$ is uniform with respect to the frequency $\lambda_j$.

Another remark is that we have very little knowledge about the constants  $c_{\cV}$ and $c_V$ (equivalently, very poor lower bounds for them). In particular, when $\cV$ is small, $c_{\cV}$ is in general much smaller than the volume of $\cV$. The consequence is that the measure $\mu_{sc}$ has no reason to be absolutely continuous w.r.t. $\mu_{L}$ (a property which would force $\mu_{sc}$ to be equal to $\mu_L$), but it could be a fractal measure, with full support. 

\subsubsection{A Fractal Uncertainty Principle}

The proof of this delocalization theorem uses similar tools as for the entropic lower bound. In order to prove a lower bound on $\mu_{sc}(\cV)$ one first smoothens the characteristic function $\bbbone_{\cV}$ into a smooth function $a_1\in C^\infty_c(T^*M)$ with similar properties as in the Definition~\ref{def:partition}. Considering eigenstates $(u_j=u_{\hbar_j})_{\hbar\to 0}$ our goal is to show the existence of a constant $C>0$ (depending only on $a_1$ but not on the eigenstates $u_\hbar$)  such that 
$$
\|u_\hbar\| \leq C \| A_1 u_\hbar\|\,.
$$
We will then complete the cutoff $a_1$ by a second cutoff $a_2\in C^\infty_c(T^*M)$, such that 
$$
a_1 + a_2 =1 \quad\text{near }S^*M\,.
$$
As above, we quantize those two cutoffs into operators $A_k=\Oph(a_k)$, such that 
$$
A_1 u_\hbar + A_2 u_\hbar = u_\hbar +\cO(\hbar^\infty).
$$
And we refine this two-member partition to define the refined operators $A_{\balpha}$ as in \eqref{e:Abalpha}. Note that each word $\balpha$ is composed only of symbols $1$ and $2$. Notice also that, if $\cV$ is a small subset of $S^*M$, then its complement $S^*M\setminus \cV$ covers most of $S^*M$, and so will the support of $a_2$. 

The quantum partition property \eqref{e:qu-partition} induces the same one for the refined partitions: for any time $n$, 
$$
u_\hbar = \sum_{|\balpha|=n} A_{\balpha}\,u_\hbar\,.
$$
As above, it will be necessary to consider times $2T_E> n > T_E$, for which the operators $A_{\balpha}$ are no more pseudodifferential operators.
The idea now will be to fix a small threshold $f\in (0,1)$, split the sum between "good" words $\balpha$ containing a fraction at least $f$ of symbols $1$, and the remaining "bad" words containing mostly $2$. This splitting allows to group the terms in the above sum into "good" and "bad" sums. 
The sum over "good" terms is bounded above by:
\begin{equation}\label{e:good}
\|\sum_{\text{good}\ \balpha} A_{\balpha} u_\hbar \| \leq \frac{C}{f} \|A_1 u_\hbar\| + o(1)\|u_\hbar\| \,.
\end{equation}
On the other hand, the "bad" terms are only controlled by the norms of the corresponding operators:
$$
\|\sum_{\text{bad}\ \balpha} A_{\balpha} u_\hbar \| \leq \sum_{\text{bad}\ \balpha} \|A_{\balpha} \|\,.
$$
The heart of the proof consists in obtaining a uniform upper bound for all the norms of the operators $A_{\balpha}$ (no matter whether they are good or bad), as long as the time $n$ is large enough. Let us treat the example of the operator $A_{\overline{2}}=A_{22\cdots 2}$. 
We remind that the dispersive estimate \eqref{e:disp} had been obtained under the condition that all elements $\Omega_k$ are sufficiently small. In the present case, the set $\Omega_2= S^*M\setminus \Omega_1 $ is not small, so we cannot directly apply the above dispersive estimate. 
We can of course partition $\Omega_2$ into small subsets $\Omega_2=\bigsqcup_{k=2}^K \Omega'_k$ and similarly with the smoothened functions $a_2=\sum_{k=2}^K a'_k$. Using this finer partition would yield similar dispersive bounds as in \eqref{e:disp} for the individual operators $A'_{\balpha}$. However, using the triangle inequality to estimate the norm of the resummed operator  $A_{\overline{2}} = \sum_{\alpha} A'_{\balpha}$ is useless: we would get
$$
\|A_{\overline{2}}\|\leq C\#\{|\balpha|=n\}\,\hbar^{-d/2}\,e^{-\Lambda n/2}\,,
$$ 
and the RHS will typically explose exponentially when $n\to\infty$, instead of decaying. 

Instead, one needs to study the {\it geometry} of the support of the function $a_{\overline{2}}$, or rather, of its symmetrized version $\tilde a_{\overline{2}}=a_{\overline{2}}\circ\Phi^{-n/2}$ as defined in \eqref{e:ta}. This function can be naturally split into
$$
\tilde a_{\overline{2}} = a^+_{\overline{2}}\,a^-_{\overline{2}}\,,
$$
where $a^+_{\overline{2}}$ (resp. $a^-_{\overline{2}}$) results from the future (resp. past) evolution, and each of them has length $n/2$. Similarly, the operator
$$
\tilde A_{\overline{2}} = A^+_{\overline{2}}\,A^-_{\overline{2}}\,.
$$
We will take $n$ as long as $2T_E$. The operator $A^+_{\overline{2}}$ is of length $\lesssim T_E$, so its shifts $U_\hbar^{n/4} A^+_{\overline{2}}U_\hbar^{-n/4}$ is a good pseudodifferential operator, with norm $\sim 1$. The same is the case for the "past operator" $A^-_{\overline{2}}$. 
But where are these operators microlocalized? 

\begin{figure}
\begin{center}
\includegraphics[width=.7\tw]{
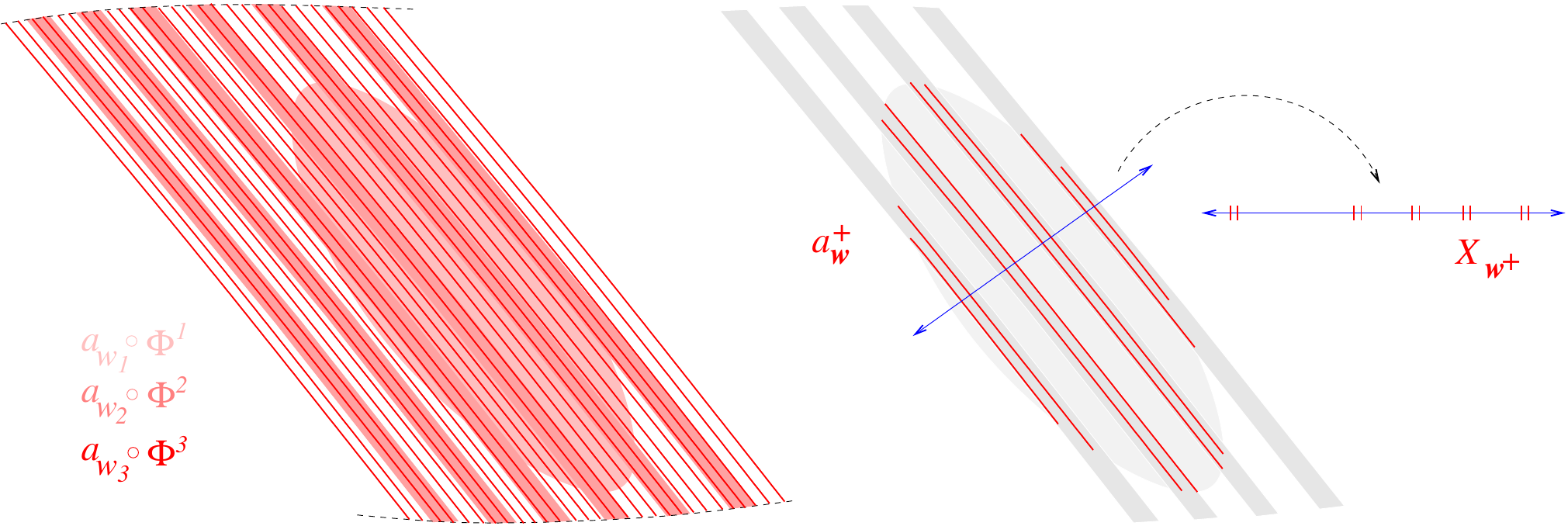}
\caption{Left: we plot the supports of $a_2\circ\Phi^t$ for several $t>0$ (different intensities of red): when $t$ grows, they align more and more with the stable manifolds $W^s$ (see also Fig.~\ref{f:evol-a1}). On the right plot, we show in red the intersection between these supports, namely the support of $\supp (a^+_{\overline{2}})$. The blue interval transverse to the stable directions features the fractal structure of this support. }\label{f:fractal1}
\end{center}
\end{figure}

\begin{figure}
\begin{center}
\includegraphics[width=.7\tw]{
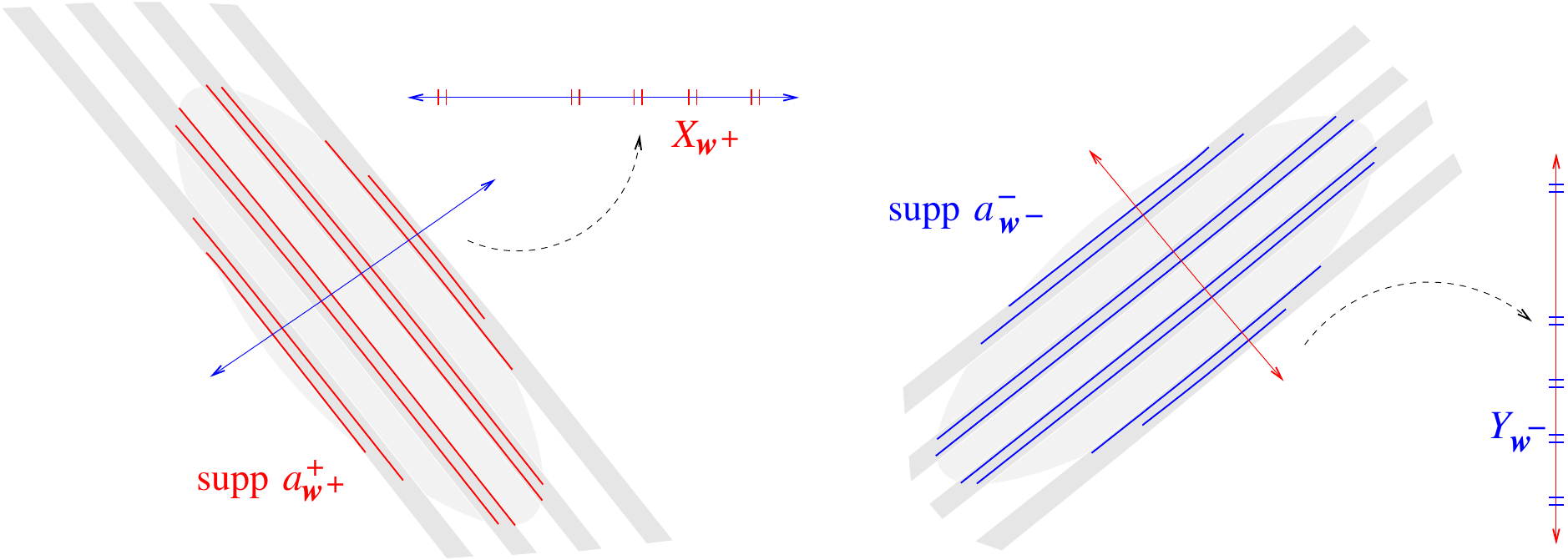}\hspace{.7cm}
\includegraphics[width=.25\tw]{
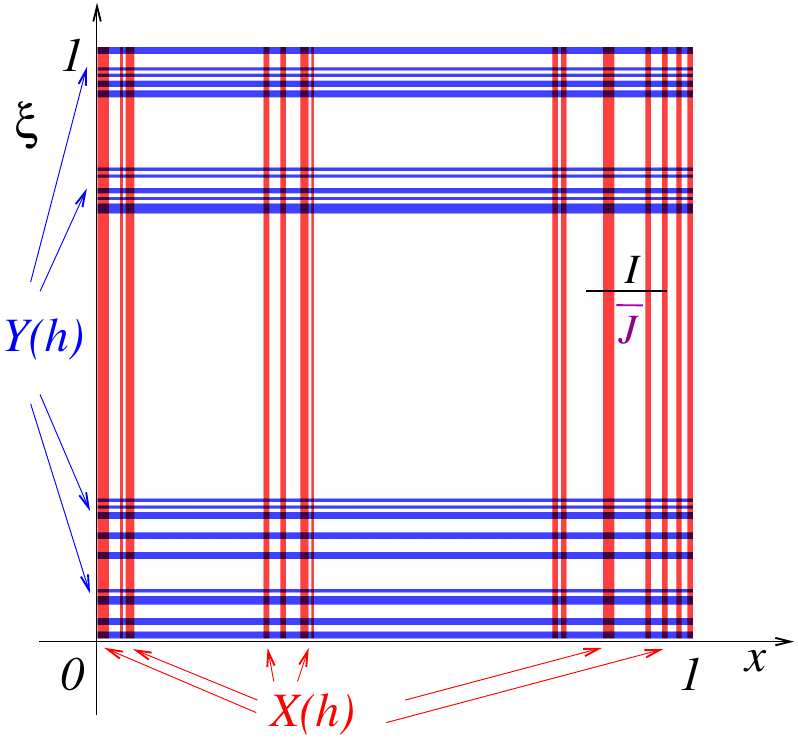}\label{f:FUP}
\caption{Left: Supports of $a^+_{\overline{2}}$ (left) and $a^-_{\overline{2}}$ (right), and the representation of their fractal structure transversely to $W^s$ (resp. $W^u$). Right: a phase space representation of the 1-dimensional FUP. }\label{f:fractal-2}
\end{center}
\end{figure}

The analysis of the function $a^+_{\overline{2}}=a_2\circ\Phi^{n/2}\cdots a_2\circ\Phi^1\,a_2$ shows that it is supported in the union of thin neighbourhoods of many stable manifolds $W^s$ (here "thin" can be as thin as $\sim \hbar$). Importantly, this support has a {\it fractal structure} transversely to the $W^s$, meaning that it is {\it porous} (has holes) on all scales between $\hbar$ and $1$ (see Figure~\ref{f:fractal1}).
The past function $a^-_{\overline{2}}$ has similar properties but along the unstable manifolds $W^u$. Its support is also porous transversely to $W^u$, on scales $\hbar$ to $1$, see Fig.~\ref{f:fractal-2}.
This fractal structure of the functions $a^{\pm}_{\overline{2}}$ now allows to invoke a recent Fractal Uncertainty Principle first stated by Dyatlov-Zahl \cite{DZ16}, and 
then proved by Bourgain-Dyatlov \cite{BD18}. This result of 1-dimensional harmonic analysis is stated as follows. 
\begin{theorem}[Fractal Uncertainty Principle]\cite{DZ16,BD18}
Let subset $X^+\subset [0,1]$ and $Y^-$ be two fractal subsets from scales $1$ to $\hbar$. The operator on $L^2(\IR)$ consisting in projecting on $X^+$, after projecting in $\hbar$-Fourier space on $Y^-\subset [0,1]$, admits the following norm estimate:\\
there exists $\beta>0$ (depending on the fractal properties of $X^+,Y^-$) such that, for $\hbar$ small enough,
$$
\|  \bbbone_{X^+}\,\cF_\hbar^*\,\bbbone_{Y^-}\cF_\hbar\|\leq \hbar^\beta\,.
$$
\end{theorem}
This estimate shows that no state $\psi_\hbar$ can be essentially localized in $X^+$, with its Fourier transform $\cF_\hbar\psi_\hbar$ being essentially localized in $X^-$. In this sense, it is an uncertainty principle. But as opposed to the usual Heisenberg uncertainty principle, it does not just rely on the sizes (lengths) of the supports of $\psi_\hbar$ and $\cF_\hbar \psi_\hbar$, but also on their fractal geometry. It is hence a subtler estimate than Heisenberg's principle.
The phase space picture of this double projection operation is indicated in Figure~\ref{f:fractal-2}.

\begin{figure}\begin{center}
\includegraphics[width=.9\tw]{
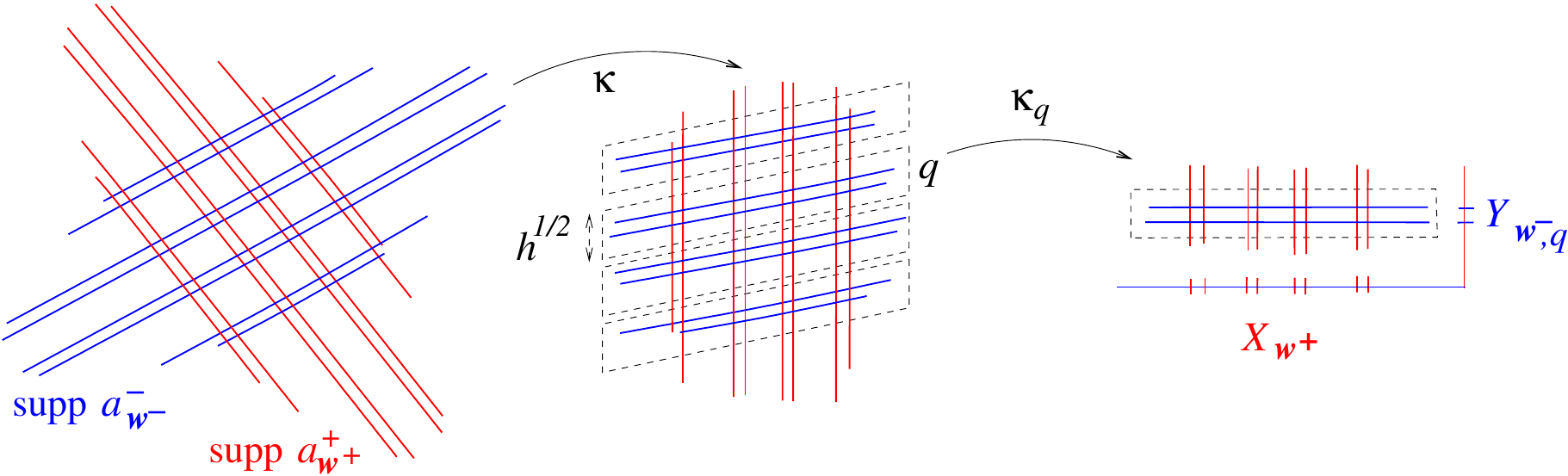}\label{f:straighten}
\caption{Successive straightenings of the supports of $a^+_{\overline{2}}$ and $a^-_{\overline{2}}$: after a first global straightening of the support of $a^+_{\overline{2}}$ through a symplectomorphism $\kappa$, one may straighten individual thin strips through symplectomorphisms $\kappa_q$.}
\end{center}
\end{figure}

How to use this FUP in our context? 
As suggested by Fig.~\ref{f:fractal-2}, the idea would be to straighten the supports of $a^\pm_{\overline{2}}$ to make them a union of vertical (resp. horizontal) thin rectangles. If this straightening could be performed through a local symplectomorphism $\kappa$, then a quantization $U_\hbar(\kappa)$ of $\kappa$, applied to the operators $A^\pm_{\overline{2}}$ would have a similar effect, namely the conjugated operators $U_\hbar(\kappa)^*A^\pm_{\overline{2}}U_\hbar(\kappa)$ would be microlocalized on a union of vertical (resp. horizontal) thin rectangles, so that one could apply the FUP to these operators, to prove that
$$
\| (U_\hbar(\kappa)^*A^+_{\overline{2}}U_\hbar(\kappa))\,(U_\hbar(\kappa)^*A^-_{\overline{2}}U_\hbar(\kappa))\| = \cO(\hbar^\beta)\,.
$$
In reality, such a global and simultaneous straightening of the supports of $a^+_{\overline{2}}$ and $a^-_{\overline{2}}$ is not possible in general. One may straighten the support of $a^+_{\overline{2}}$ into a union of vertical rectangles through a symplectomorphism $\kappa$, but the support of $a^-_{\overline{2}}\circ\kappa^{-1} $ is then not made of parallel rectangles. One may split the latter support into a union of tilted strips, and then straighten each strip independently of the other ones through adapted symplectomorphisms $\kappa_q$, so that  $a^+_{\overline{2}}\circ\kappa^{-1} \circ\kappa_q^{-1}$ remains a union of vertical thin rectangles, while $a^-_{\overline{2}}\circ\kappa^{-1} \circ\kappa_q^{-1}$ is a union of thin horizontal rectangles (see Fig.~\ref{f:straighten}). One can then apply the FUP to the two operators $U_\hbar(\kappa_q\circ\kappa)^*A^\pm_{\overline{2}}U_\hbar(\kappa_q\circ\kappa)$, because they respectively resemble $\bbbone_{X^+}$ and $\cF_\hbar^*\,\bbbone_{Y^-}\cF_\hbar$ in the theorem. 
Finally, one  realizes that the different strips are sufficiently separated from each other in phase space, so that the corresponding operators are ``orthogonal" to each other. The bounds obtained for each individual strip then also applies to their sum, so one gets the required bound when $n\sim 2T_E$:
$$
\| A^+_{\overline{2}} \| = \| A^+_{\overline{2}}\,A^-_{\overline{2}}\| = \cO(\hbar^\beta).
$$
Here we explained how to obtain this nontrivial norm bound for the operator $A_{\overline{2}}$. The same proof actually applies to any symbolic word $A_{\balpha}$ of same length $n$. 

Finally, we want to sum those bounds over all the bad words $\balpha$. Choosing a parameter $0<\vareps $ smaller than $\beta$, we may select the threshold $f$ small enough, such that the number of ``bad words" grows at most like $\hbar^{-\vareps}$. The triangle inequality then ensures that 
$$
\|\sum_{\text{bad}\ \balpha}A_{\balpha}\|\leq \hbar^{\beta-\vareps}=o(1)\quad \text{when}\ \hbar\to 0\,.
$$
Summing this contribution to the ``good" one in \eqref{e:good}, we end up with 
$$
\|u_\hbar\|\leq  \frac{C}{f} \|A_1 u_\hbar\| + o(1)\|u_\hbar\|\,,
$$
which is the required bound.

In this proof we see that the dependence of the constant $C/f$ w.r.to the set $\cV\subset S^*M$ essentially comes from the exponent $\beta$ in the FUP, which conditions the choice of the threshold $f$. The dependence of the exponent $\beta$ w.r.to the fractal structure of $X^+, Y^-$ is poorly known.
\hfill$\square$

\subsection{Failure of QUE for discrete time toy models}
As explained after Theorem~\ref{thm:QE}, there exists discrete time toy models with strongly chaotic dynamics (Anosov diffeomorphisms) which admit a form of quantization into ``quantum maps", for which the same questions of eigenstate delocalization can be addressed. One can show a quantum ergodicity theorem \cite{BDB96,MOK05,DENW06}, and therefore wonder whether QUE also holds in this enlarged context. 
For a specific family of quantum maps, namely the quantization of Arnold's ``cat map"  on the 2-dimensional torus phase space, discovered by Hannay and Berry \cite{HB80}, one can show lower bounds for the entropy of semiclassical measures as in Theorem~\ref{thm:entropy} \cite{Bro10}. However, 
due to the strong algebraic properties of this family, one could exhibit counterexamples to QUE, namely sequences of eigenmodes corresponding to semiclassical measures $\mu_{sc}$ different from the Liouville measure \cite{FNDB03}, namely the convex combination of $\mu_L$ and of measures supported on individual periodic orbits. The entropies of these $\mu_{sc}$ can be as small as the threshold value \eqref{e:Riv10}, that is half of the value for $\mu_{L}$.
On the opposite, analogues of Lindenstrauss's ``Hecke" eigenmodes for this system satisfy an arithmetic form of QUE \cite{KR05}.
The full delocalization of semiclassical measures for the 2-dimensional cat maps was proved by Schwartz in \cite{Schw24}.

Other counterexamples to QUE were found for higher dimensional Anosov diffeomorphisms \cite{Kel10}, with semiclassical measures supported on submanifolds of the phase space. Entropic lower bounds were obtained for higher dimensional cat maps in \cite{Riv10b}. Constraints on the delocalization of semiclassical measures in higher dimensional cat maps were obtained by Dyatlov and Jezequel in \cite{DJ24}: they showed that the partially localized semiclassical measures found by Kelmer are not always present, depending on the arithmetic properties of the classical map. 

Let us remark that those counterexamples to QUE are quite different from the one presented in section~\ref{s:Hassell}: those diffeomorphisms are uniformly hyperbolic, hence they do not admit any family of neutral periodic orbits. 

\section{Conclusions and perspectives}

In these chapter we have described various results pertaining to the macroscopic distribution of eigenmodes of the Laplacian on manifolds or domains with a chaotic geodesic flow. The main focus has been on the Quantum Ergodicity, a now classical result intiated by Schnirelman in the 1970s, which described the huge majority of the eigenmodes. This result only requires the ergodicity of the classical flow, and its can be accomodated to a large variety of systems, including systems with singularities. Several questions remains open, like the precise statistical properties and rate of decay of the matrix elements $(\la u_j,\Op_{\hbar_j}(\chi-\mu_L(\chi))u_j\ra)_j$. Conjectures have been proposed for this statistics, connected with Random Matrix Theory, but few rigorous results are available. In the case of a general manifold of negative curvature, Zelditch proved a logarithmic decay \cite{Zel94} for the variance.  Luo-Sarnak proved an essentially sharp rate of decay of the quantum variance for the Hecke eigenstates on the (finite volume yet noncompact) modular surface in \cite{LS04}. A similar decay was proved by Kurlberg-Rudnick for the Hecke eigenstates of the quantum cat map \cite{KR05}: in those works, it is shown that most matrix elements decay like $\sqrt{\hbar}$. On the opposite, Schubert \cite{Schu08} shows that for certain non-Hecke eigenbases of the quantum cat map, the quantum variance may decay much slowlier. 

Other aspects of Quantum Ergodicity have not been accounted for, like the possibility to test the equidistribution in semiclassically small regions (say, in phase space balls of radii $r(\hbar)\to 0$, or on spatial balls of similar radii $r(\hbar)\to 0$. Such results have been obtained initially for Anosov systems by Han \cite{Han15} and Hezari-Rivi\`ere \cite{HR16}, where the balls decay like some $|\log\hbar|^{-\alpha}$. Similar small scale quantum ergodicity has been obtained by Han for quantum cat map eigenstates \cite{Han18}. 

Quantum ergodicity in space (but not in phase space) has been shown for several nonergodic systems, like the Laplacian on rational polygons \cite{MR12}. In this case, the space equidistribution comes from the projection properties of the Lagrangian surface carrying the eigenmodes. Spatial small-scale quantum ergodicity was proved for toral eigenfunctions \cite{LR17}.  

For most chaotic systems, it is believed that the spectrum of the corresponding quantum operator (say, the Laplacian) is nondegenerate: the eigenbasis is then unique. However, this spectral simplicity is usually not proven, but only conjectured. This is why the requirement to take Hecke eigenstates of arithmetic surfaces remains necessary to obtain QUE, eventhough it is believed that there is no other eigenbasis. For the quantum cat map, the situation is very different: the spectrum may be highly degenerate, leaving the possibility to select eigemodes of various types: this is in such a case of "maximal multiplicity" that one was able to construct non-QUE eigenmodes in \cite{FNDB03}. Having such a high freedom of choice for eigenbases leads one to consider quantum (unique) ergodicity for {\it random eigenbases}. This path has been pursued by the author together with N.Schwartz \cite{NS25} for the quantum cat map. 
More generally, one can relax the eigenmode condition and enlarge the study to random quasimodes of specific spectral widths, for systems with or without classical chaotic behaviour. This path of research has been put forward in particular by S.Zelditch see e.g. \cite{Zel92, Zel14, BL13, Map13, Han17, HT20}. 
\smallskip

Concerning the approach to QUE for deterministic modes, let us mention that the entropy bounds of Thm~\ref{thm:entropy} have been recently generalized by Eswarathasan \cite{Esw25} to quasimodes of the Laplacian of spectral widths $\sim \alpha\frac{\lambda}{\log\lambda}$: he shows how the entropy lower bounds decrease when one increases the parameter $\alpha$ (for $\alpha$ large enough, one can construct quasimodes localized along a single closed geodesic). 

Most recent results on the approach to QUE are attempting to extend the use of the Fractal Uncertainty Principle to higher dimensional manifolds or systems. Dyatlov-Jezequel \cite{DJ24} managed to use the 1-dimensional FUP to address certain families of higher-dimensional cat maps, namely the ones featuring one large simple eigenvalue of the classical map, implying a strong unstable direction. Upon an irreducibility of the characteristic polynomial of the classical map, they prove full equidistribution of all semiclassical measures. In the more general case where the characteristic polynomial is not irreducible over the rationals, the support of a semiclassical measure can be a strict subset of $\IT^{2n}$, but it must contain some invariant subtorus. The counterexamples of Kelmer \cite{Kel10} are covered by those results.

Similar results were obtained by Athreya-Dyatlov-Miller \cite{ADM25} for the Laplacian on compact quotients of complex hyperbolic spaces: those are compact manifolds of dimension $2n$, of negative curvature, possessing one stronger unstable direction, like in the preceding example of the cat map. In that case, the authors show that for any semiclassical measure $\mu_{sc}$ of the Laplacian, the support of $\mu_{sc}$ must contain some $S^*\Sigma$, where $\Sigma$ is a compact immersed totally geodesic complex submanifold of $M$. For most such manifolds, $\Sigma$ is necessarily the full manifold $M$, so one recovers a full delocalization of $\mu_{sc}$. 

In these two examples of systems, the presence of a stronger Lyapunov exponent allowed to use the 1-dimensional FUP, ajusted to the fast stable and unstable directions. If this stronger Lyapunov exponent does not exist, it is necessary to use a higher-dimensional FUP. The generalization of the 1-dimensional FUP to higher dimensions is not obvious, there are simple counterexamples to a naive generalize statements. Cohen \cite{Coh25} proved a statement of FUP in higher-dimension, which could be used by Kim and Miller \cite{KM25} to show a form of delocalization for hyperbolic manifolds (that is, with constant negative curvature) in higher dimension. The result is similar to the one of Athreya et al. on complex hyperbolic quotients:  they prove that the support of a semiclassical measure necessarily contains $S^*\Sigma$ for $\Sigma$ some compact immersed totally geodesic submanifold.

Both examples of higher dimensional manifolds are locally symmetric spaces, meaning that all points all the manifold locally look the same. In this situation, the stable/unstable subspaces depend smoothly on the base point, which allows to use specific tools. Relaxing this homogeneity seems to represent a major hurdle: the stable/unstable subspaces will then depend H\"older-continuously on the base point, which will pose serious difficulties when applying semiclassical methods. Another difficult point will be the description of the fractal sets appearing in higher dimension. On surfaces, we showed that the fractality boiled down to fractal subsets $X^+$, $Y^-$ of the unit interval. In higher dimensional, the analogues of $X^+$, $Y^-$ will be fractal subsets of $\IR^n$ for $n\geq 2$: the zoology of such fractal sets is much richer and more complicated than in 1 dimension. A good understanding of the geometry of these sets will be necessary to apply any form of FUP.
\smallskip

Finally, let me emphasize that these notes only described the macroscopic (sometimes mesoscopic, in the small scale quantum ergodicity results) behaviour of the eigenmodes. A different focus would concern the {\it microscopic} behaviour of these modes, that is the structure of their fluctuations at the wavelength scale $\hbar_j=\lambda_j^{-1}$. This is a completely different focus, and mathematical results are quite scarce. It was conjectured by Berry in 1977 \cite{Be77} that at this scale, the eigenmodes of chaotic manifolds look like random combinations of monochromatic waves of various directions. This so-called Random Wave Model has itself attracted a lot of attention from probabilists and hamonic analysts during the last 20 years, who studied at length its statistical properties. Yet, showing that this model indeed applies to chaotic eigenmodes still seems out of reach.

\begin{ack}[Acknowledgments]

During the write-up of this chapter, the author has been partially supported by the, grant ANR-20-CE40-0017 (project ``Al\'eatoire, Dynamique et Spectre" (ADYCT)). 
\end{ack}


\bibliographystyle{JHEP}%
\bibliography{Nonnen}

\end{document}